\documentclass[sigconf, 9pt]{acmart}

\AtBeginDocument{%
  \providecommand\BibTeX{{%
    \normalfont B\kern-0.5em{\scshape i\kern-0.25em b}\kern-0.8em\TeX}}}

\setcopyright{acmlicensed}

\copyrightyear{2024}
\acmYear{2024}
\setcopyright{rightsretained}
\acmConference[ICCAD '24]{IEEE/ACM International Conference on Computer-Aided Design}{October 27--31, 2024}{Newark, NJ, USA}
\acmBooktitle{IEEE/ACM International Conference on Computer-Aided Design (ICCAD '24), October 27--31, 2024, Newark, NJ, USA}
\acmDOI{10.1145/3676536.3676762}
\acmISBN{979-8-4007-1077-3/24/10}

\usepackage{titlesec}
\usepackage{enumitem}
\usepackage{microtype}
\usepackage{graphicx}
\usepackage{booktabs}
\usepackage{tikz}
\usepackage{amsmath}
\usepackage{mathtools}
\usepackage{subcaption}
\usepackage{amsthm}
\usepackage{textcomp}
\usepackage{amsmath}
\DeclareMathOperator*{\argmax}{arg\,max}

\usepackage[capitalize,noabbrev]{cleveref}

\theoremstyle{plain}

\theoremstyle{definition}

\theoremstyle{remark}

\usepackage[textsize=tiny]{todonotes}
\usepackage{multirow}
\usepackage{graphicx}
\usepackage{caption}
\usepackage{longtable}

\begin{document}

\newcommand{\pgftextcircled}[1]{
    \setbox0=\hbox{#1}%
    \dimen0\wd0%
    \divide\dimen0 by 2%
    \begin{tikzpicture}[baseline=(a.base)]%
        \useasboundingbox (-\the\dimen0,0pt) rectangle (\the\dimen0,1pt);
        \node[circle,draw,outer sep=0pt,inner sep=0.1ex] (a) {#1};
    \end{tikzpicture}
}


\title{MapTune: Advancing ASIC Technology Mapping via Reinforcement Learning Guided Library Tuning}


\author{Mingju Liu}
\affiliation{%
  \institution{University of Maryland, College Park}
   \country{Maryland, USA}
  }
\email{mliu9867@umd.edu}
\author{Daniel Robinson}
\affiliation{%
  \institution{Massachusetts Institute of Technology}
   \country{Massachusetts, USA}
  }
\email{daniel_r@mit.edu}
\author{Yingjie Li}
\affiliation{%
  \institution{University of Maryland, College Park}
    \country{Maryland, USA}
  }
\email{yingjiel@umd.edu}
\author{Cunxi Yu}
\affiliation{%
  \institution{University of Maryland, College Park}
   \country{Maryland, USA}
  }
\email{cunxiyu@umd.edu}

\begin{abstract}
    Technology mapping involves mapping logical circuits to a library of cells. Traditionally, the full technology library is used, leading to a large search space and potential overhead. Motivated by randomly sampled technology mapping case studies, we propose MapTune framework that addresses this challenge by utilizing reinforcement learning to make design-specific choices during cell selection. By learning from the environment, MapTune refines the cell selection process, resulting in a reduced search space and potentially improved mapping quality. 

    The effectiveness of MapTune is evaluated on a wide range of benchmarks, different technology libraries and technology mappers. The experimental results demonstrate that MapTune achieves higher mapping accuracy and reducing delay/area across diverse circuit designs, technology libraries and mappers. The paper also discusses the Pareto-Optimal exploration and confirms the perpetual delay-area trade-off.
    Conducted on benchmark suites ISCAS 85/89, ITC/ISCAS 99, VTR8.0 and EPFL benchmarks, the post-technology mapping and post-sizing quality-of-results (QoR) have been significantly improved, with average Area-Delay Product (ADP) improvement of 22.54\% among all different exploration settings in MapTune. The improvements are consistently remained for four different technologies (7nm, 45nm, 130nm, and 180 nm) and two different mappers.

\end{abstract}


\maketitle

\section{Introduction}

Targeted specialization of functionality in hardware has become arguably the best means for enabling improved compute performance and energy efficiency. However, as the complexity of modern hardware systems explodes, fast and effective hardware explorations are hard to achieve due to the lack of guarantee in the existing in electronic design automation (EDA) toolflow. Several major limitations prevent practical hardware explorations \cite{kapre2015intime,ziegler2017ibm,yu2018end}. First, as the hardware design and technology advance, the design space of modern EDA tools has increased dramatically. Besides, evaluating a given design point is extremely time-consuming, such that only a very small sub-space of the large design space can be explored. Last but not least, while the initialization of design space exploration is important for the final convergence, it is difficult to initialize the search for unseen designs effectively.


Recent years have seen increasing employment of decision intelligence in EDA, which aims to reduce the manual efforts and boost the design closure process in modern toolflows \cite{kapre2015intime,ustun2019lamda,liu2013-ml-hls,pasandi2019approximate,ziegler2017ibm,li2016efficient,DBLP:conf/dac/YuXM18,hosny2019drills,ma2019high,yu2020decision,yu2020flowtune}. For example, various of machine learning (ML) techniques have been used to automatically configure the tool configurations of industrial FPGA toolflow \cite{kapre2015intime,ustun2019lamda,liu2013-ml-hls,schafer2019high,liu2019accelerating,neto2022flowtune} and ASIC toolflow \cite{ziegler2017ibm,li2016efficient,DBLP:conf/dac/YuXM18,hosny2019drills,}. These works focus on end-to-end tool parameter space exploration, which are guided by ML models trained based on either offline \cite{ziegler2017ibm} or online datasets \cite{kapre2015intime,ustun2019lamda}. Moreover, exploring the sequence of synthesis transformations (also called synthesis flow) in EDA has been studied in an iterative training-exploration fashion through Convolutional Neural Networks (CNNs) \cite{DBLP:conf/dac/YuXM18} and reinforcement learning \cite{hosny2019drills}. While the design quality is very sensitive to the sequence of transformations \cite{DBLP:conf/dac/YuXM18}, these approaches are able to learn a sequential decision making strategy to achieve better quality-of-results \cite{DBLP:conf/dac/YuXM18,hosny2019drills}. Moreover, \cite{neto2022flowtune,liu2024cbtune} demonstrate the effectiveness of lightweight Multi-Arm Bandit (MAB) models in identifying the optimal synthesis flow and it achieves a balance between exploring and exploiting arms through multiple trials to maximize overall payoffs. In addition, neural network based image classification and image construction techniques have been leveraged in placement and route (PnR), in order to accelerate design closure in the physical design stage \cite{xie2018routenet,yang2018gan,xu2019wellgan, 
mirhoseini2020chip, 
yu2019painting,yang2019deepattern,zhongdeep2020dac}.  
As the design of digital circuits continues to grow in complexity, technology mapping process faces an increasingly large search space due to the vast number of cells contained in modern technology libraries. Utilizing the entire technology library as an input for technology mapping can result in excessive runtime and sub-optimal results, which has been demonstrated with our comprehensive case studies using 7nm ASAP library \cite{clark2016asap7,neto2021read,neto2021slap}.

To address this issue, we argue that carefully calibrated partial technology libraries that contain a subset of cells have been proposed to mitigate the search space and reduce runtime. However, the selection of an optimal subset of cells requires significant expertise and experience to carefully consider the design goals, target technology, and characteristics of each cell. This process is known as cell selection and represents a significant challenge for EDA research. In addition, the optimal selection of cells process could result in a new area-delay trade-off. Thus, identifying whether there exists new performance Pareto frontier from the technology mapping during the exploration stage is of great interest. 

Thus, this paper presents a novel Reinforcement Learning guided sampling framework to explore the design space of partially selected technology library in optimizing and exploring the technology mapping performance, without changing the mapping algorithms, namely MapTune. The main contributions of this work are summarized as follows:
\begin{itemize}[leftmargin=1mm]
\item We present a comprehensive case study utilizing the 7nm ASAP library \cite{clark2016asap7} to showcase the performance implications of various partially random-sampled libraries. Our study reveals substantial variations in delay (up to 40\%) and area (up to 60\%) across selected designs.
\item In this paper, we introduce a novel cell selection framework, MapTune, based on Multi-Armed Bandit (MAB) and Q-Learning, seamlessly integrated within the ABC framework \cite{brayton2010abc}. This framework facilitates effective library tuning for technology mapping, wherein timing evaluations are performed during the post-sizing stage using Static Timing Analysis (STA) techniques.
\item We evaluate MapTune framework using designs from five distinct benchmark suites: ISCAS 85/89 \cite{brglez1989combinational}, ITC/ISCAS 99 \cite{basto2000first}, VTR8.0 \cite{luu2014vtr}, and EPFL benchmarks \cite{soeken2018epfl} mapped on four different libraries: 7nm ASAP library \cite{clark2016asap7}, FreePDK45  45nm libary \cite{freepdk45}, SKYWATER 130nm library \cite{sky130google}, GlobalFoundries 180nm MCU library \cite{gf180google}. The results demonstrate average Area-Dealy Product (ADP) improvements of 22.54\% by solely tuning the libraries.
\item 
MapTune will be released as open-source project in the integration of ABC \cite{brayton2010abc} framework at \url{https://github.com/Yu-Maryland/MapTune}.
\end{itemize}

\section{Background}

\subsection{Technology Mapping and Library}

Technology mapping is a critical phase in the logic synthesis process, converting high-level circuit descriptions, such as those at the Register Transfer Level (RTL), into technology-specific gate-level netlists, particularly for Application-Specific Integrated Circuits (ASIC) designs. It involves selecting appropriate cells from an EDA library to realize a circuit in a chosen technology, effectively bridging high-level design with physical implementation. This step not only follows logic optimization but is also essential for optimizing Power, Performance, and Area (PPA), significantly influencing the cost, performance, and manufacturability of a circuit by determining the optimal gates and their interconnections.

Various algorithms have been developed to address the technology mapping problem, crucial in logic synthesis for ASIC design. These include tree-based approaches \cite{marwedel1993tree,cong1994flowmap}, which focus on mapping trees or sub-trees to specific gates, and Directed Acyclic Graph (DAG)-based methods \cite{mishchenko2006improvements,mishchenko2005technology,mishchenko:2006-dag} that consider the entire circuit topology for enhanced optimization. Additionally, genetic algorithms \cite{kommu1993gafpga}, inspired by natural selection, use a population of solutions evolved over time to find optimal or near-optimal configurations for technology mapping. Recently, ML approaches have also been involved in improving the technology mapping process, either by correlating the technology-independent representation to technology-dependent PPAs (prediction models) \cite{neto2021read,li2024boolgebra}, or by directly optimizing the technology mapping algorithms \cite{neto2021slap}. These techniques aim to optimize PPA by minimizing gate count and interconnections, reducing power consumption, and meeting timing constraints. Thus, technology mapping is essential for producing efficient, cost-effective, and high-performance circuits in modern electronic systems.


{Technology libraries are critical for technology mapping, as they provide predefined gates and components optimized for specific fabrication processes. Consequently, significant efforts have been made to optimize or generate standard cell libraries to improve the PPAs in the design flow \cite{ren2021nvcell}. However, there has been no effort to analyze the impact of these libraries on the technology mapping procedure. Specifically, in this work, we observe and demonstrate a counter-intuitive finding: \textit{a partially selected set of cells from the full technology library might significantly improve the technology mapping PPAs.} The main intuition behind this is that the complexity and algorithmic space of technology mapping algorithms increase as the number of cells in the libraries increase, while most practical technology mappers in academic and industrial toolflows are heavily heuristic-based. Our comprehensive case studies in Section \ref{sec:case_study} will first discuss the impact of specific cell selection within a full library on the technology mapping PPA results using the ABC framework. 
}

\subsection{Learning-based techniques in EDA}

Learning-based techniques have found widespread application across various aspects of the EDA area, including synthesis, placement and routing, and design space exploration for different design stages. Numerous studies have achieved substantial success in addressing placement problems using reinforcement learning \cite{mirhoseini2020chip,elgammal2021rlplace,yu2018developing,ren2023machine}. 
Additionally, various studies such as \cite{liang2020drc}, \cite{8942063}, 
\cite{alawieh2020high},
have explored different learning techniques, including Convolutional Neural Networks (CNN) and Generative Adversarial Networks (GAN), during the routing phase. 
Another notable application is Design Space Exploration (DSE) for logic synthesis. The optimization of quality-of-results (QoR) in logic synthesis often requires extensive tuning runtime, making efficient DSE a challenging task due to the exponential number of potential design permutations. In response to this challenge, \cite{haaswijk2018deep, yu2018developing} have applied GAN and CNN techniques to automate design space exploration and synthesis design flow. 

\section{MapTune Case Studies}\label{sec:case_study}

Technology mapping plays a crucial role in the logic synthesis process within the domain of EDA. An excessively large technology mapping library can impose significant pressure on the exploration of design space and make the search for optimal gate selection challenging. Therefore, reducing the library size in a reasonable manner presents opportunities for optimizing technology mapping. Thus, to explore the impact of library sampling size on technology mapping performance, we conduct a comprehensive case study that involves random sampling from the 7nm ASAP library \cite{clark2016asap7}. 

\begin{figure*}
    \centering
    \begin{subfigure}[b]{0.32\textwidth}
        \includegraphics[width=0.95\textwidth]{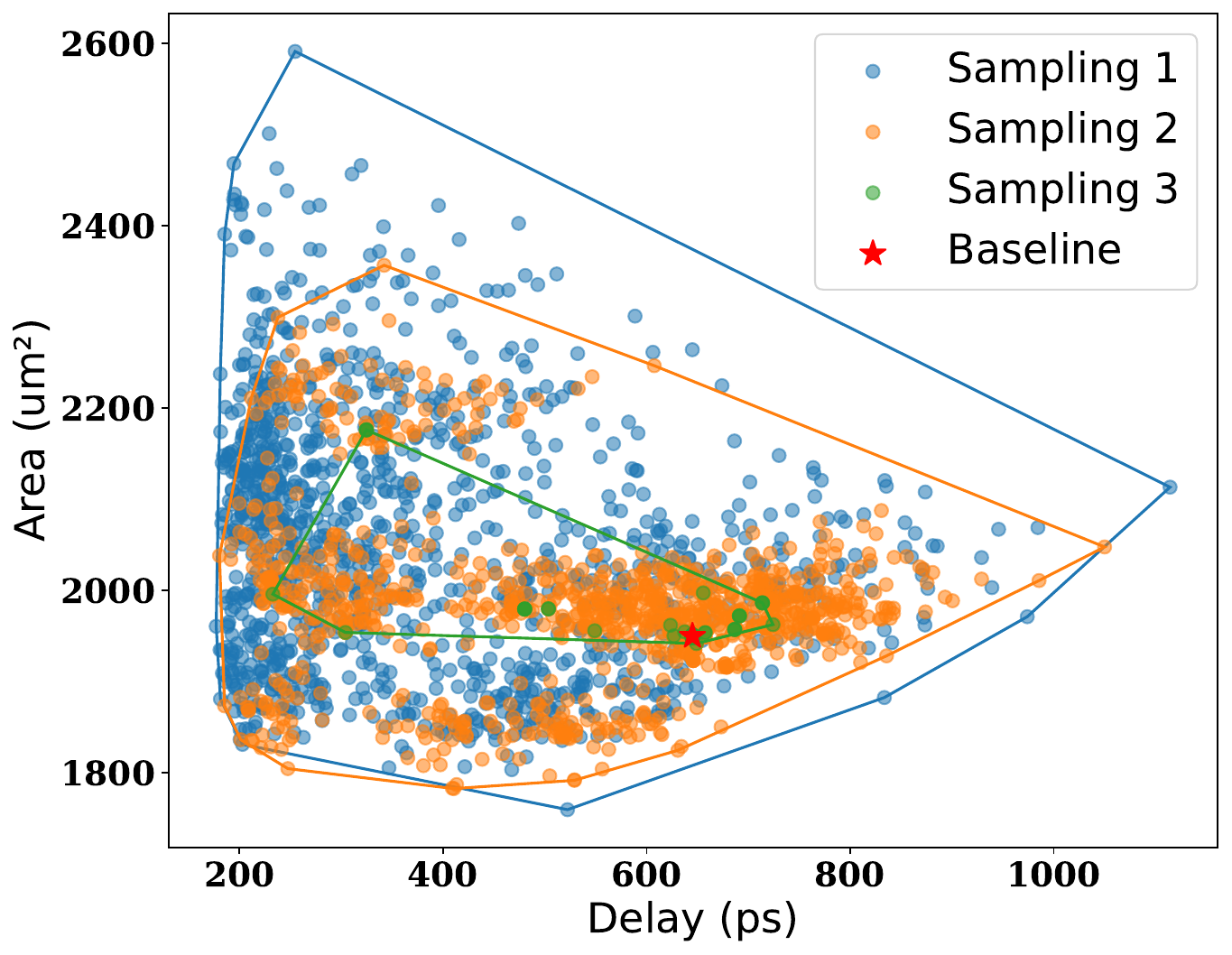}
        \caption{s13207}
        \label{fig:s13207}
    \end{subfigure}
    \hfill
    \begin{subfigure}[b]{0.32\textwidth}
        \includegraphics[width=0.95\textwidth]{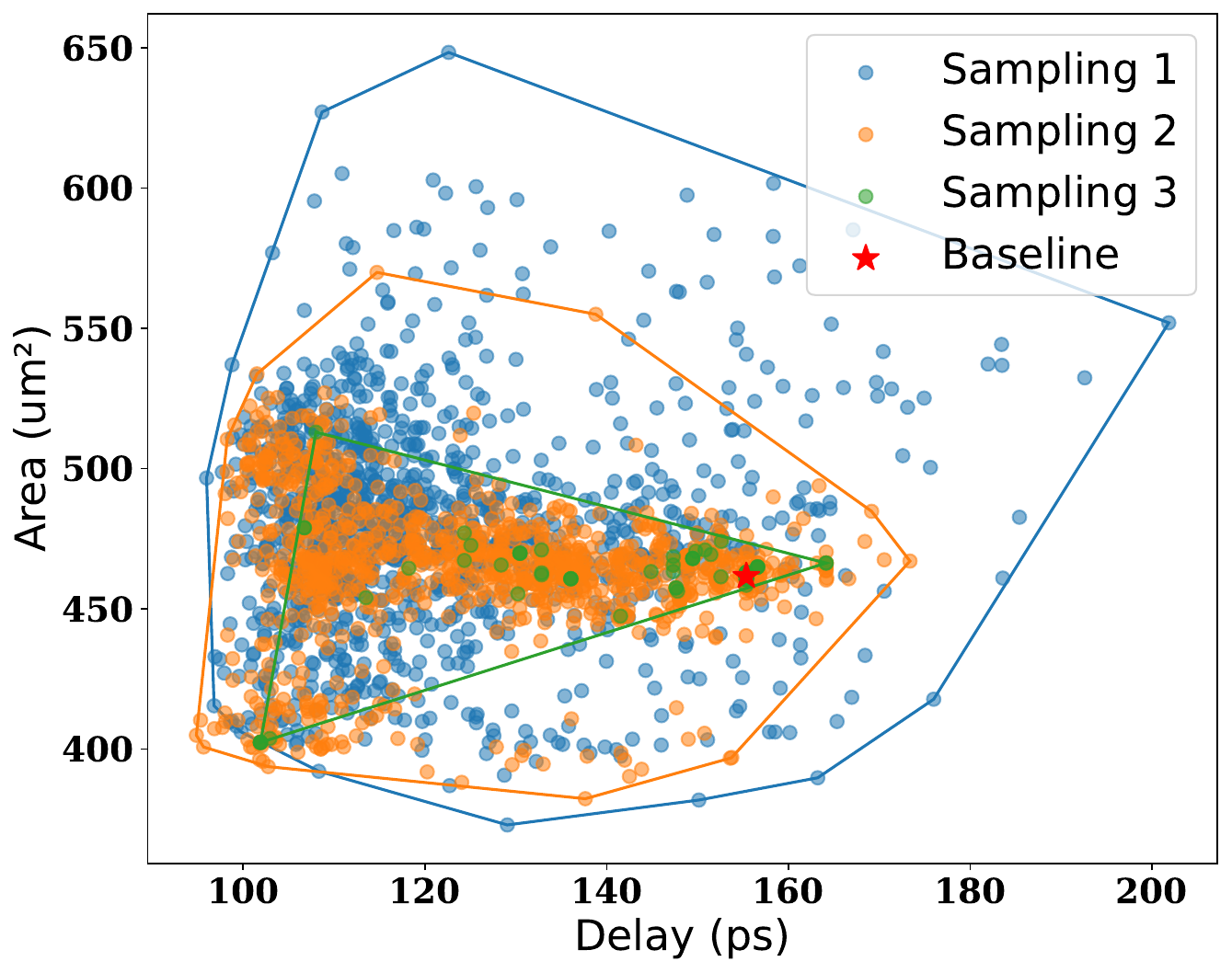}
        \caption{c2670}
        \label{fig:c2670}
    \end{subfigure}
    \hfill
    \begin{subfigure}[b]{0.32\textwidth}
        \includegraphics[width=0.95\textwidth]{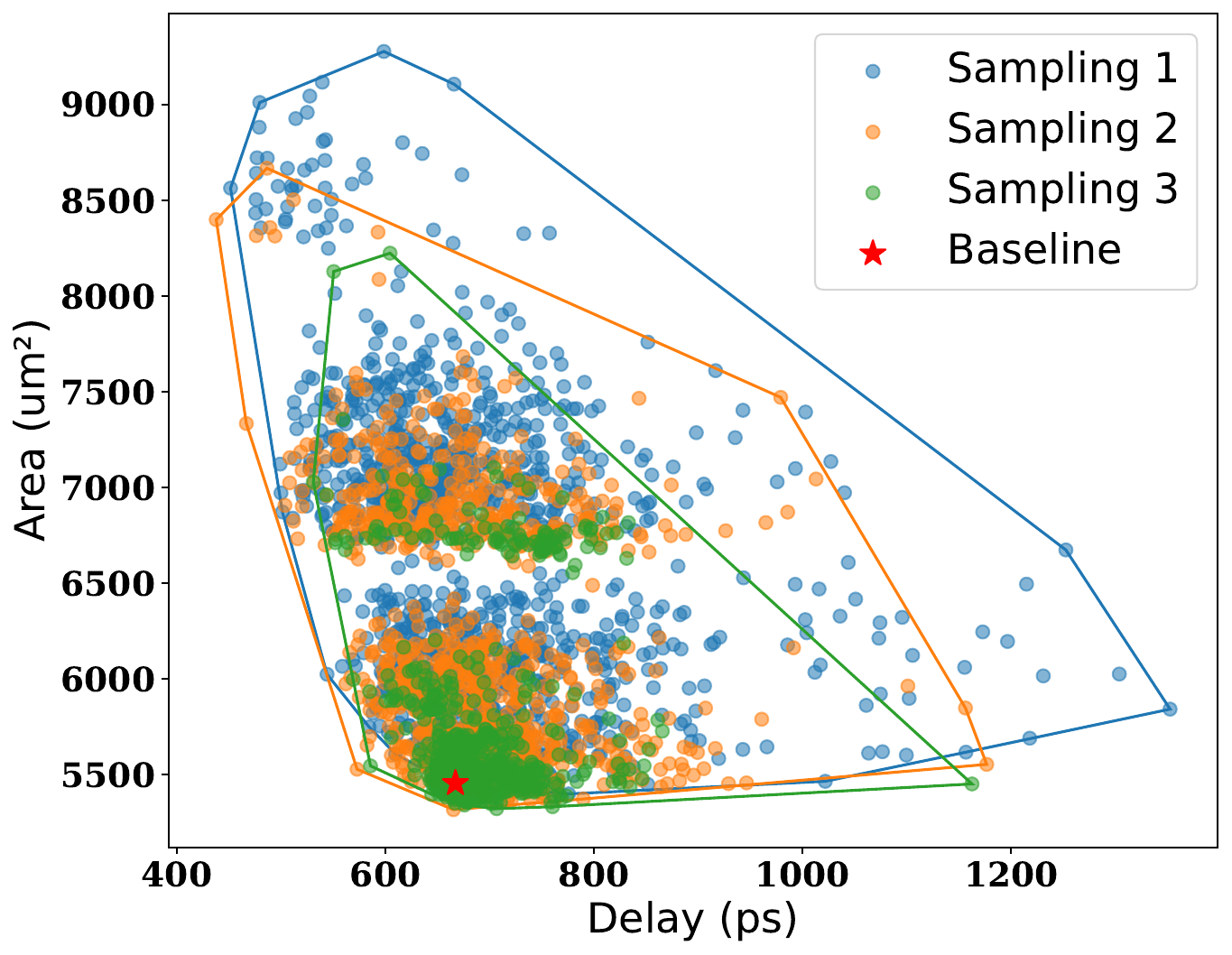}
        \caption{b20}
        \label{fig:b20}
    \end{subfigure}
    \caption{Technology mapping results of selected designs: Baseline: All 161 cells of \texttt{ASAP7} library; Sampling 1: Randomly sampling 75 - 100 cells; Sampling 2: Randomly sampling 100 - 125 cells; Sampling 3: Randomly sampling 125 - 150 cells.}
    \label{fig:random_sample_3}
\end{figure*}

In particular, we use the complete 7nm ASAP technology library with 161 cells as the baseline. This baseline approach represents the conventional method that is widely employed in the industry. 
Then, we randomly sample the technology library space using three different sampling ranges: 1) 75 -- 100 cells, 2) 100 -- 125 cells, and 3) 125 -- 150 cells, denoted as Sampling 1, Sampling 2, Sampling 3 in Figure \ref{fig:random_sample_3}, respectively. Note that all tests with different random-sampled partial libraries are conducted on synthesis framework ABC \cite{mishchenko2007abc} with mapping, gate sizing, and STA timing analysis commands\footnote{\texttt{read library.lib;map;topo;upsize;dnsize;stime}}. The case study results are presented in Figure \ref{fig:random_sample_3}.

By undertaking this case study, we aimed to investigate optimization opportunities through various library sampling sizes. Our analysis primarily focused on design-specific scenarios, aiming at providing valuable insights into the selection of an appropriate sampling size for efficient technology mapping. Based on the experimental results summarized in Figure \ref{fig:random_sample_3}, three critical observations have been summarized as follows: 

\textbf{Observation 1 -- The sampled partial libraries has significant impacts on QoR in the post-sizing stage.} 
It is evident that the results of the baseline, indicated by the red star, fall within the distribution range of QoR obtained from the sampled partial libraries. Consider design $\texttt{s13207}$ illustrated in Figure \ref{fig:s13207} as an example. In terms of area, $\sim$30\% of the results from the sampled libraries tend to outperform the baseline, while in terms of delay, $\sim$60\% exhibit better performance. By reducing the number of available components, technology mapping can focus on a refined library of cells that aligns precisely with the specific design requirements. This approach has demonstrated promising results, facilitating fine-grained optimizations and improved performance. 
 
\textbf{Observation 2 -- QoR distributions vary significantly with different sampling sizes.} 
We discover that larger sampling sizes tend to generate more clustered results, while smaller sampling sizes reveal greater potential for optimization. Consider design $\texttt{c2670}$ illustrated in Figure \ref{fig:c2670}. The results obtained from the three different sampling sizes exhibits various improvements in both delay and area optimization compared to the baseline. Notably, the smallest sampling range (75 - 100 cells), represented by the blue dots, reveals a wider distribution of QoR than the two larger sampling sizes. This disparity suggests that certain components may be underutilized when larger sampling sizes are employed, thereby missing out on the opportunity to achieve better quality-of-results. This observation provides us with insight into the importance of reaching a balance between exploitation and exploration, which is a key consideration in developing our framework.

\textbf{Observation 3 -- Technology mapping on partially sampled libraries likely outperforms the baseline but not for all designs.}
From the two aforementioned observations, we can draw an initial conclusion that partially sampled libraries for technology mapping are likely to achieve better results compared to using the full library. However, it is important to note that there are exceptions that contradict this conclusion. For example, when examining design $\texttt{b20}$ in Figure \ref{fig:b20}, we observe that the baseline point falls within the lower left corner of the range of results obtained from the three sampled libraries. This suggests that narrowing down the library size for this specific design does not entirely benefit technology mapping. Additionally, we must consider another potential drawback of smaller sampling sizes: while smaller sizes offer greater potential for improved results, there is a risk of failed mapping due to the reduced number of available components, which limits the ability to find a suitable mapping solution for certain designs. Therefore, it is crucial to carefully evaluate the trade-off between optimization objectives and the probability of successful mapping when determining the appropriate sampling size. 

Through this case study, we have highlighted the impact of partially sampled library on the the potential for achieving better results, and the trade-off between optimization and successful mapping. Our observations underscore the importance of tuning the technology mapping library to meet design-specific requirements. 
Furthermore, this study motivates us to explore a new framework to address this specific problem iteratively by leveraging learning-based techniques to hold promise for advancing technology mapping in design-specific applications.

\section{Approach}\label{section:approach}

\vspace{-3mm}
\begin{table}[h]
\centering
\caption{Notations of MapTune formulation.}
\resizebox{\columnwidth}{!}{
\begin{tabular}{|c|l|}
\hline
\textbf{Notation} & \textbf{Description} \\
\hline
$\mathcal{L}$ & Set of all cell variants in the Library file \\
$\mathcal{N}$ & Number of cells variants in the Library file \\
$\mathcal{A}$ & Action space with a discrete and finite set of actions \\
$a^i$ & Action that select cell variant $i$\\
$\mathcal{S}$ & A state indicates which actions have been taken/which cell variants have been chosen \\
$ADP_{\mathcal{S}}$ & Normalized Area-Delay Product under current state $\mathcal{S}$\\
$p_{a^i}$ & Probability of cell variant $i$ minimizing Area-Delay Product \\

\hline
\end{tabular}}
\label{tab:notation}
\end{table}
\vspace{-3mm}
\subsection{Formulation of MapTune}

\begin{figure}
    \centering
    \includegraphics[width=\columnwidth]{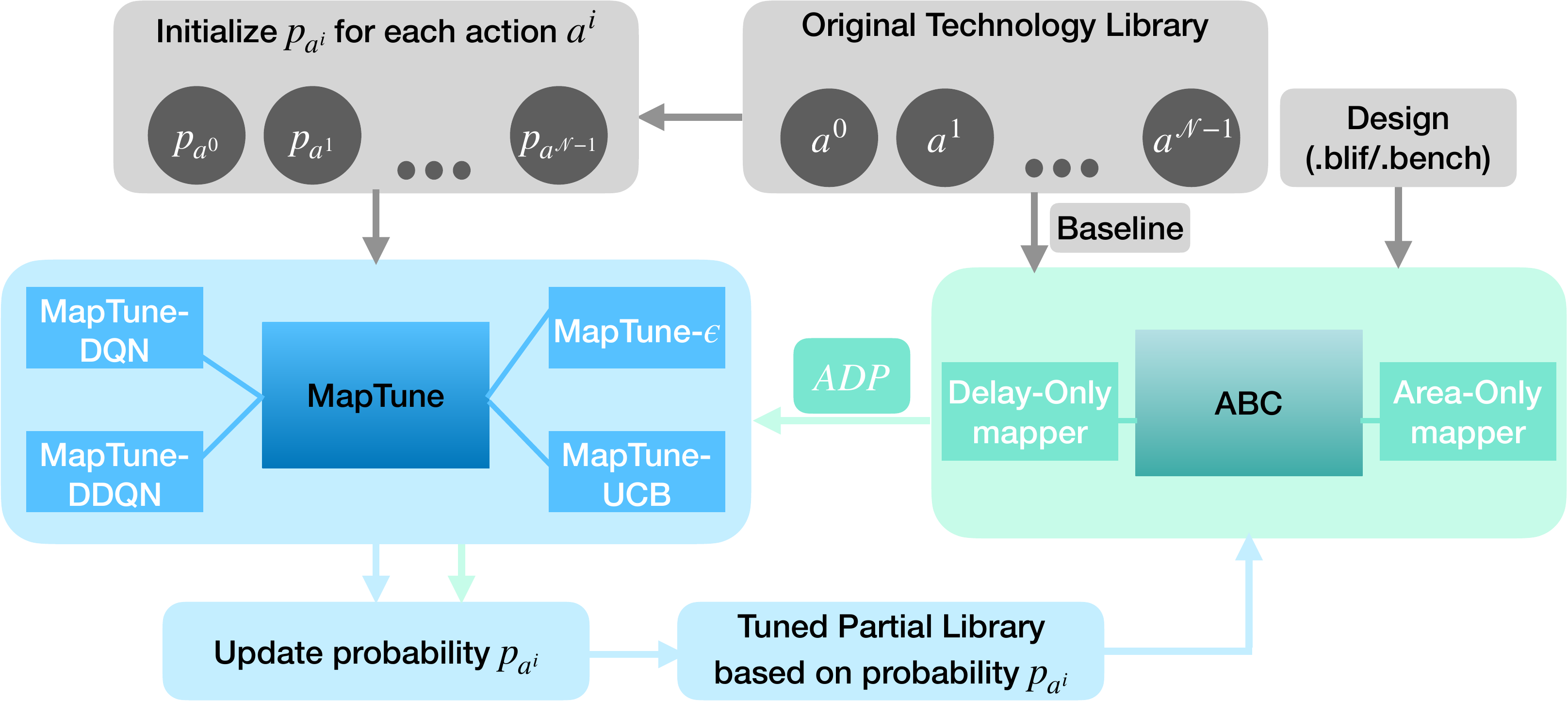}
    \caption{MapTune Framework Overview. Termination is based on the user-defined number of iterations for $p$ update.}
    \label{fig:flow}
\end{figure}

An overview of the MapTune framework is shown in Figure \ref{fig:flow}. MapTune is dedicated to addressing a cell selection problem with a pool of $\mathcal{N}$ candidates, denoted as $\mathcal{L}$ (LibSet). Each candidate in $\mathcal{L}$ is associated with two performance metrics: $Delay$ and $Area$. In this work, we use the Area-Delay Product ($ADP$) as a single metric to assess overall circuit efficiency. By definition, $ADP = Delay \times Area$. In each decision iteration, a subset of $n$ candidates is selected from $\mathcal{L}$, forming a potential solution. Each candidate $i \in \mathcal{L}$ corresponds to a binary decision variable $S_i \in \{0,1\}$. A solution is then represented by a multi-hot encoded vector $\mathcal{S} \in \{0,1\}^{\mathcal{N}}$, where $\mathcal{S} = [S_0, S_1, \ldots, S_{\mathcal{N}-1}]$ and $\sum_{i=0}^{\mathcal{N}-1}S_i = n$. Each distinct $\mathcal{S}$ represents a unique combination of candidates, each associated with a specific reward.

Here are the essential model formulation configurations:

\textbf{Action Space:} The action space $\mathcal{A}$ consists of a discrete and finite set of actions, where each action corresponds to selecting a specific cell to form a subset of cells from the technology library, subsequently utilized for the technology mapping of a design. The cardinality of this action space is equal to $\mathcal{N}$, denoting the total number of unique cells in the initial technology library. Here, we define each action as $a^i: S_i = 1, i \in [0, \mathcal{N}-1], a^i \in \mathcal{A}$, where taking action $a^i$ means selecting cell $i$ from the original library.

\textbf{State:} During each iteration of forming a subset of cells from the original library with all candidate cells, the state $\mathcal{S}$ represents the current data collection condition, where $\mathcal{S} = [S_0, S_1, \ldots, S_{\mathcal{N}-1}]$. And when $\sum_{i=0}^{\mathcal{N}-1}S_i = n$, where $n$ is the required subset size, it refers to a complete state.

\textbf{Reward:} The reward function corresponds to the cell selections made by the agent. It is defined as $\mathcal{R}_{\mathcal{S}} = -ADP_{\mathcal{S}} = -\left(\frac{D_{\mathcal{S}}}{D_{Base}} \cdot \frac{A_{\mathcal{S}}}{A_{Base}}\right)$, where $D_{\mathcal{S}}$ and $A_{\mathcal{S}}$ are the metrics derived from the technology mapping of the design using the selected subset of cells indicated at state $\mathcal{S}$. The terms $D_{Base}$ and $A_{Base}$ represent the baseline metrics, established using all the cells in the original library for technology mapping. In this case, $ADP_{\mathcal{S}}$ is a product of normalized $D_{\mathcal{S}}$ and $A_{\mathcal{S}}$. The negative function is employed to invert the metric, encouraging maximizing the reward by minimizing $ADP$, thus optimizing both metrics concurrently.

Formally, we define the probability vector as $\boldsymbol{p} = [p_{a^0}, p_{a^1}, \ldots, \\p_{a^{\mathcal{N}-1}}]$. \textbf{The probability $p_{a^i}$ is defined as the likelihood of selecting cell $i$ for the sampled library can maximize the reward.} These vectors are updated at each decision epoch based on the observed performance metric, i.e., $ADP$. The objective is to iteratively refine $\boldsymbol{p}$ such that the probability of selecting candidates that lead to minimizing $ADP$ is maximized.

\subsection{Implementation}
In the realm of Reinforcement Learning, we choose two directions to formulate the MapTune Framework: Multi-Armed Bandit (MapTune-MAB) and Q-Learning (MapTune-Q). More specifically, our prominent MapTune-MAB algorithms are with $\epsilon$-greedy strategy \cite{kuleshov2014algorithms} and Upper Confidence Bound strategy \cite{jouini2010upper}, denoted as MapTune-$\epsilon$ and MapTune-UCB, respectively. For MapTune-Q methods, we implement and compare both the Deep Q-Network (DQN) \cite{mnih2015human} and the Double Deep Q-Network (DDQN) \cite{van2016deep}.

\subsubsection{MapTune-MAB}

As for the bandit problem settings, we refer to each cell $i \in \mathcal{L}$ in the library as an arm $i$, during each iteration, if an action $a^i$ is taken, this means arm $i$ is selected for this action.

\textbf{MapTune-$\epsilon$ Agent.} The MapTune-$\epsilon$ agent utilizes $\epsilon$-greedy to balance exploration and exploitation via a parameter $\epsilon \in [0, 1]$. This parameter dictates the probability of random action selection (exploration) vs. choosing the action with the highest probability leads to higher reward based on historical data (exploitation). Formally, the agent selects action $a^i \in \mathcal{A}$ according to the following rule:

$a^i = \left\{\begin{matrix}
\argmax_{a^i \in \mathcal{A}} p_{a^i}& \text{with parameter } 1-\epsilon \\ 
\text{a random selection } a^i \in \mathcal{A} & \text{with parameter } \epsilon
\end{matrix}\right.$

\textbf{MapTune-UCB Agent.} The MapTune-UCB agent integrates a confidence interval around the reward estimates based on historical trial data to tackle the similar exploration-exploitation problem effectively. Action selection is governed by the following formula:
\begin{equation}
    a^i = \argmax_{a^i \in \mathcal{A}} (p_{a^i} + c \sqrt{\frac{\text{log}(t)}{n_{a^i}}})
\end{equation}
where $t$ is the current iteration, $n_{a^i}$ is the number of times that action $a^i$ is taken during $t$ iterations , $c$ is the coefficient that modulates the extent of exploration.

Note that, for both MapTune-$\epsilon$ and MapTune-UCB agents, the probability vector $\boldsymbol{p}$ are updated as following: 

\begin{equation}
    p_{a^i}(t+1) = \frac{p_{a^i}(t) n_{a^i}(t) + \mathcal{R}_{\mathcal{S}}(t)}{n_a(t)}
\end{equation}
where $p_a(t)$ is the probability of action $a^i$ at iteration $t$, $n_{a^i}(t)$ is the number of times that action $a^i$ is taken during $t$ iterations. In this context, the probability of taking action $a^i$ leads to minimizing $ADP$ is an average of the obtained reward during the data trial process. Note that, we ensure every time when updating $p_{a^i}$ for the next iteration $(t+1)$ at state $\mathcal{S}(t)$, a required number of arms has been selected, so that the reward $\mathcal{R}_{\mathcal{S}}$ is influencing a subset of arms during each iteration.

\subsubsection{MapTune-Q}

Following the same action space, state and reward setup, we can implement MapTune-Q Agent similarly. Instead of influencing the probability of action decision directly from the observed reward, our prominent MapTune-Q methods facilitate the DQN to approximate the optimal state-action value function with the associated reward. Both MapTune-DQN Agent and MapTune-DDQN Agent are implemented as following by adapting the same environment settings:

\textbf{MapTune-DQN Agent.}Given a state vector $\mathcal{S}$ and action $a_i$ as the input of the DQN, the model will predict a Q-value $Q(\mathcal{S}, a^i)$. To keep the consistency, we use the same probability $p_{a^i} = Q(\mathcal{S}, a^i)$ to refer to the predicted Q-value of the taken action $a^i$. The model also derives a target Q-value, denoted as $p_{a^i}^{tar}$ through Bellman equation as follows:
\begin{equation}
    p_{a^i}^{tar} = \mathcal{R} + \gamma \max_{a^i} p_{a^i} \label{eq:bellman}
\end{equation}
where $\mathcal{R}$ is the current reward introduced by taking action $a^i$, and $\gamma$ is the discount factor to emphasize the significance of the future rewards. Note that, only after a certain number of actions have been taken reaches the required selection size, the $\mathcal{R}$ will be reflected as the actual $-ADP$ as aforementioned.

During each iteration, we use the DQN by parameter $\theta$ to calculate the Q-function and choose the action with the highest probability (i.e., $p_{a^i}$) leads to maximizing the reward. For the trainable parameters of DQN, we use Mean Squared Error (MSE) as the loss function to calculate between the predicted Q-value ($p_{a^i}$) and target Q-value ($p_{a^i}^{tar}$), then through backward propagation, the network parameters are updated hence affecting further probabilities of chosen actions. It can defined as follows:
\begin{equation}
    L_{\theta} = MSE(p_{a^i}, p_{a^i}^{tar}) 
\end{equation}

\textbf{MapTune-DDQN Agent.} To mitigate the overestimation bias in the DQN due to the maximization step in the Bellman Equation \ref{eq:bellman}, we use DDQN to optimize this process. DDQN incorporates additional network, the target DQN ($Q_{target}$), which mirrors the online DQN ($Q_{online}$) in terms of number and configurations of layers but with only periodically updated weights. When taking a state $\mathcal{S}$ and action $a^i$, both $Q_{target}$ and $Q_{online}$ will produce a Q-value denoted as, $p_{a^i} = Q_{online}(\mathcal{S},a^i)$ and $p'_{a^i} = Q_{target} (\mathcal{S}, \argmax_{a^i} p_{a^i})$, respectively. Consequently, we calculate the target Q-value $p_{a^i}^{tar}$ as following:
\begin{equation}
    p_{a^i}^{tar} = \mathcal{R} + \gamma p'_{a^i}
\end{equation}

While $Q_{online}$ update the parameters using the same settings as in MapTune-DQN Agent, the weight update of $Q_{target}$ is under a soft update rule as follows:
\begin{equation}
    \theta_{target} \leftarrow \tau \theta_{online} + (1-\tau) \theta_{target}
\end{equation}
where $\tau$ is a small coefficient that controls the rate of the update hence stabilizing the learning process.

\color{black}

\section{Results}

We demonstrate the proposed approach on designs from five benchmark suites: ISCAS 85, ISCAS 89, ITC/ISCAS 99, VTR8.0 and EPFL benchmarks, to evaluate the performance of MapTune after technology mapping and gate sizing are performed. Specifically, MapTune explores the technology libraries and evaluates it by mapping on the library using ABC with the same command as in Section \ref{sec:case_study}. 
Note that, by default, the ABC built-in \texttt{map} command is a Delay-driven mapper. To provide more robustness to this work, we also use \texttt{map -a} command in ABC which is an Area-driven mapper by default for experiments.
All experiments are conducted with an Intel\textregistered Xeon\textregistered Gold 6418H CPU and NVIDIA RTX\texttrademark A6000 GPU. 
We evaluate the mapped results with different sampling sizes in MapTune, which searches for the best achievable results. Through the observed results, we evaluate the design space exploration using MapTune, which aims to search for Pareto Frontier in area-delay trade-offs. All experiments are conducted with the 7nm ASAP library \cite{clark2016asap7}, FreePDK45  45nm libary \cite{freepdk45}, SKYWATER 130nm library \cite{sky130google}, GlobalFoundries 180nm MCU library \cite{gf180google}. For simplicity, we will refer to them as \texttt{ASAP7}, \texttt{NAN45}, \texttt{SKY130}, and \texttt{GF180} respectively.

All results presented in this section are conducted with MapTune framework given a one-hour timeout constraint. All MapTune-MAB and MapTune-Q methods are implemented with a batch size of 10. Following the approach used in the motivating case studies, MapTune is evaluated by fixing its sampling size range over the same technology library among different optimization methods. To ensure a fair comparison, sampling sizes are set as follows: 1) 45 - 135 cells; 2) 35 - 75 cells; 3) 220 - 310 cells; and 4) 40 - 130 cells for \texttt{ASAP7}, \texttt{NAN45}, \texttt{SKY130}, and \texttt{GF180} library, respectively, with a step size of 10 within the sampling range.

\subsection{Technology Mapping Results}
\begin{figure*}[htbp]
    \centering
    \begin{subfigure}[b]{0.32\textwidth}
        \includegraphics[width=0.99\textwidth]{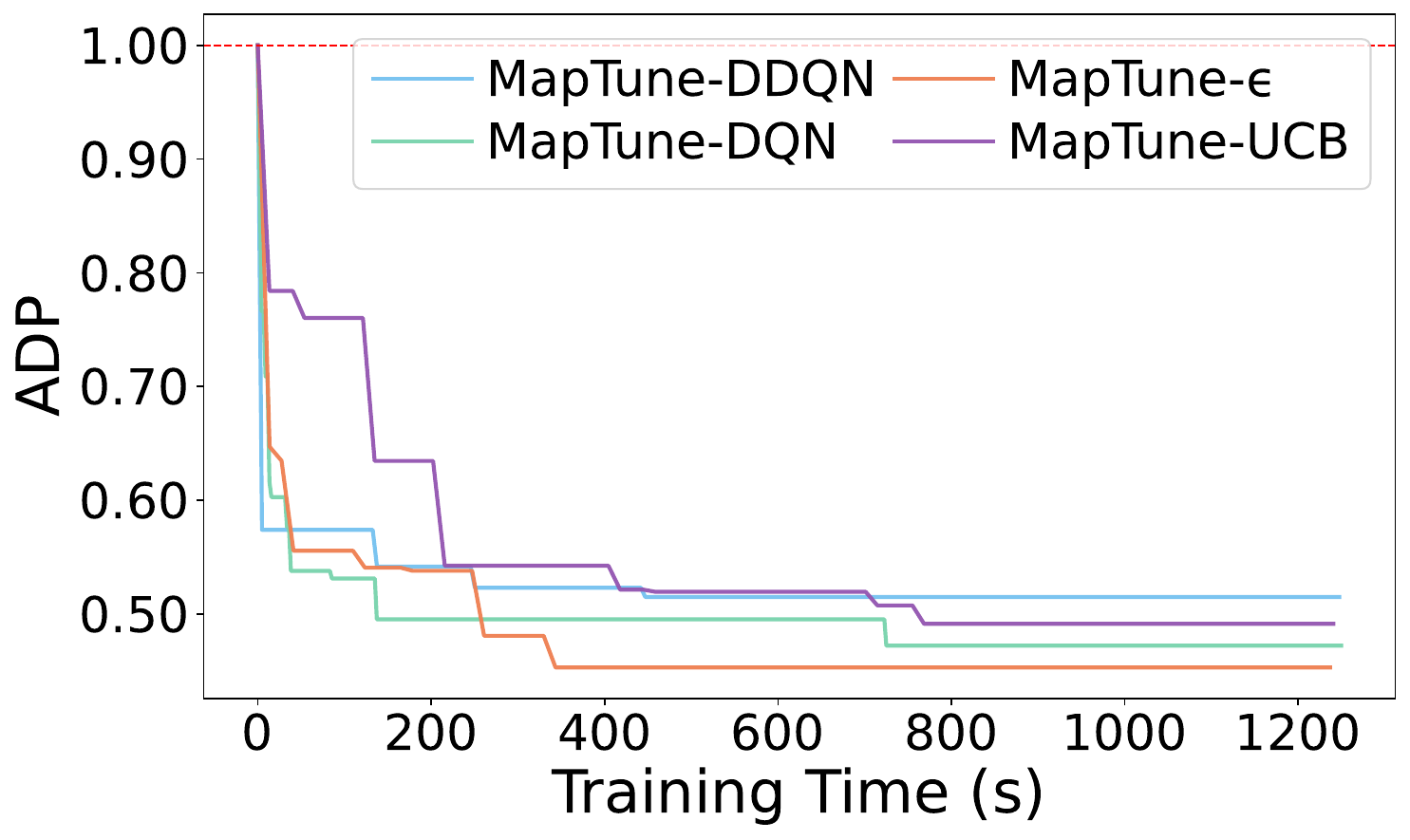}
        \caption{b14}
        \label{fig:comp7:b14}
    \end{subfigure}
    \hfill
    \begin{subfigure}[b]{0.32\textwidth}
        \includegraphics[width=0.99\textwidth]{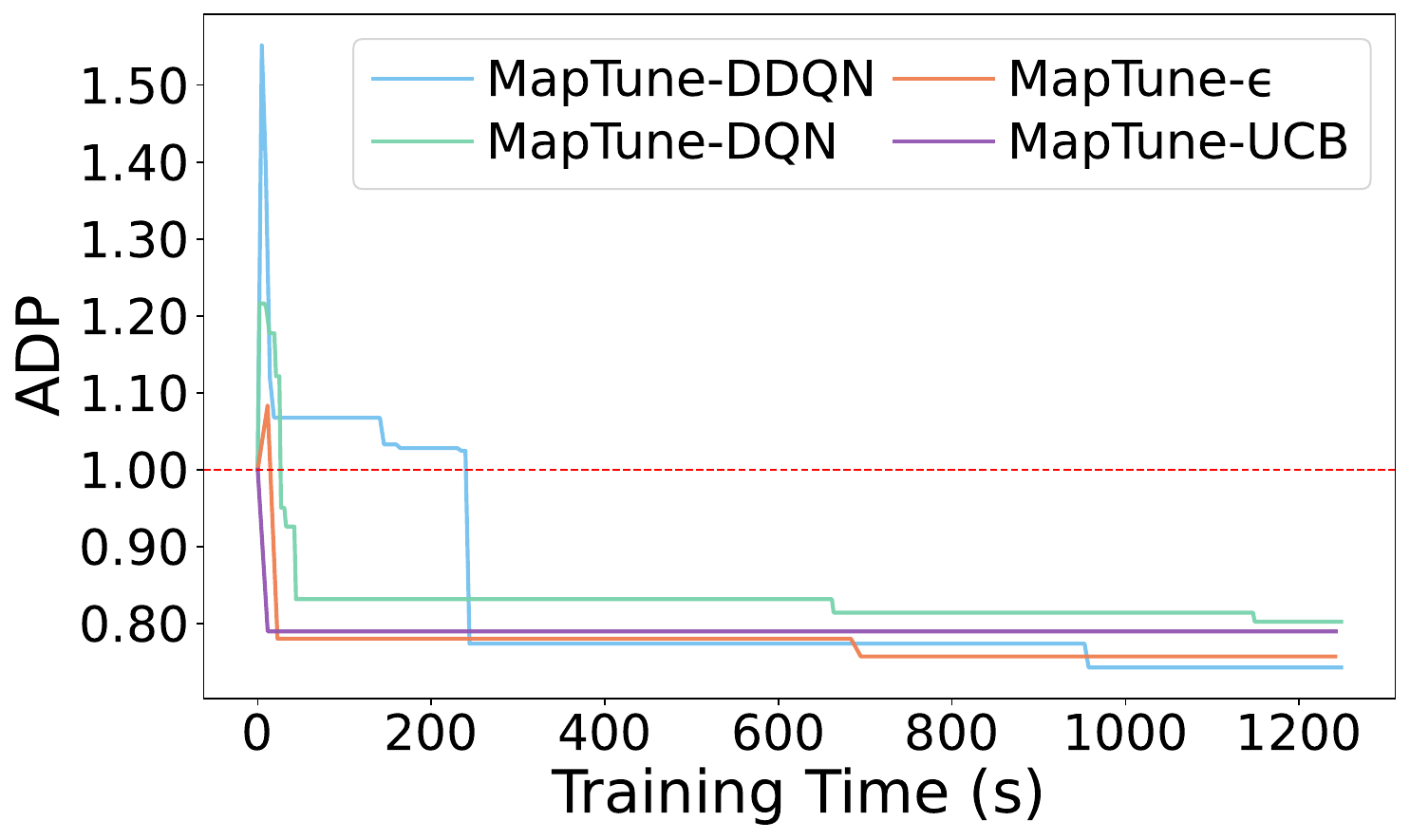}
        \caption{bar}
        \label{fig:comp7:bar}
    \end{subfigure}
    \hfill
    \begin{subfigure}[b]{0.32\textwidth}
        \includegraphics[width=0.99\textwidth]{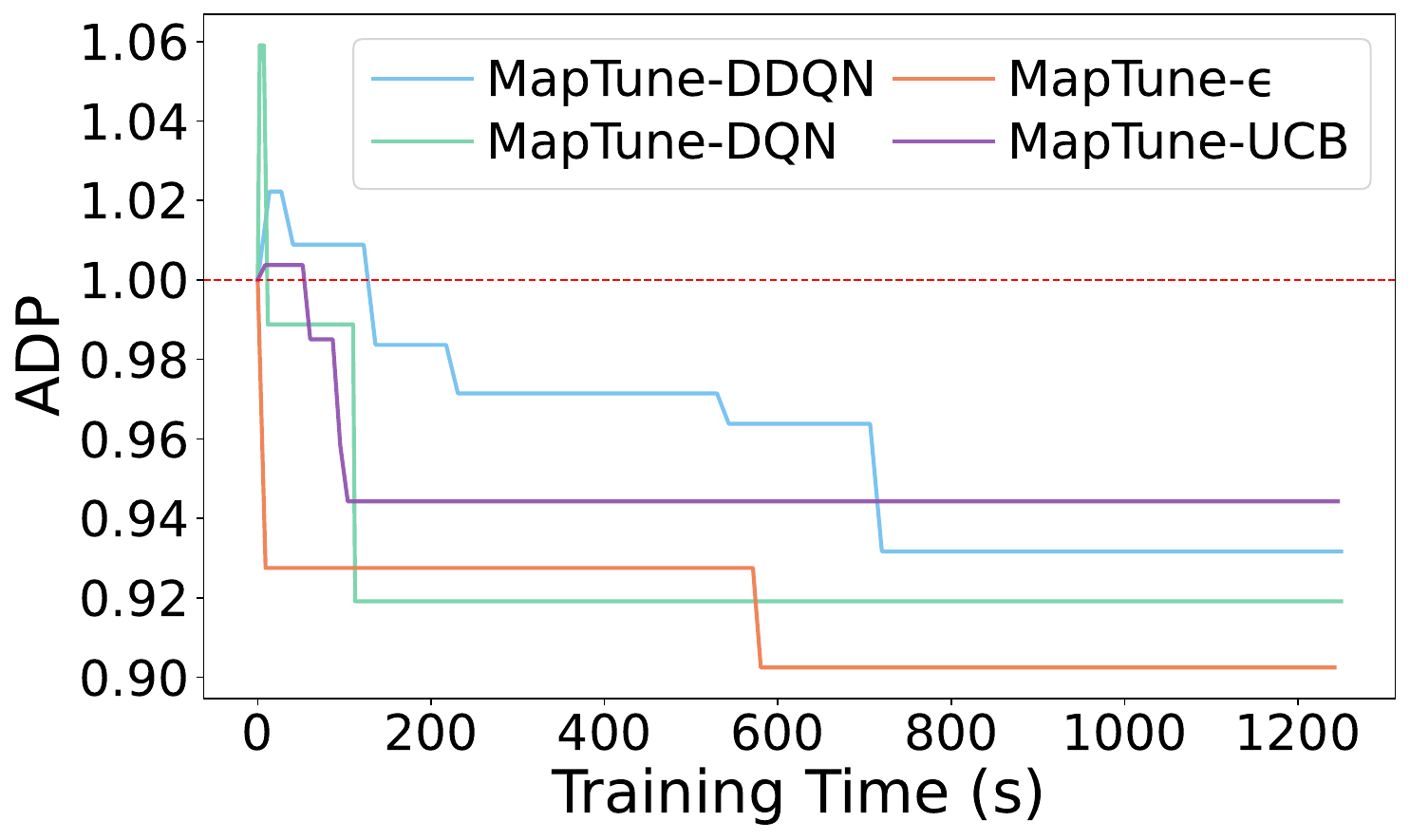}
        \caption{c1238}
        \label{fig:comp7:c1238}
    \end{subfigure}
    
    \medskip
    
    \begin{subfigure}[b]{0.32\textwidth}
        \includegraphics[width=0.99\textwidth]{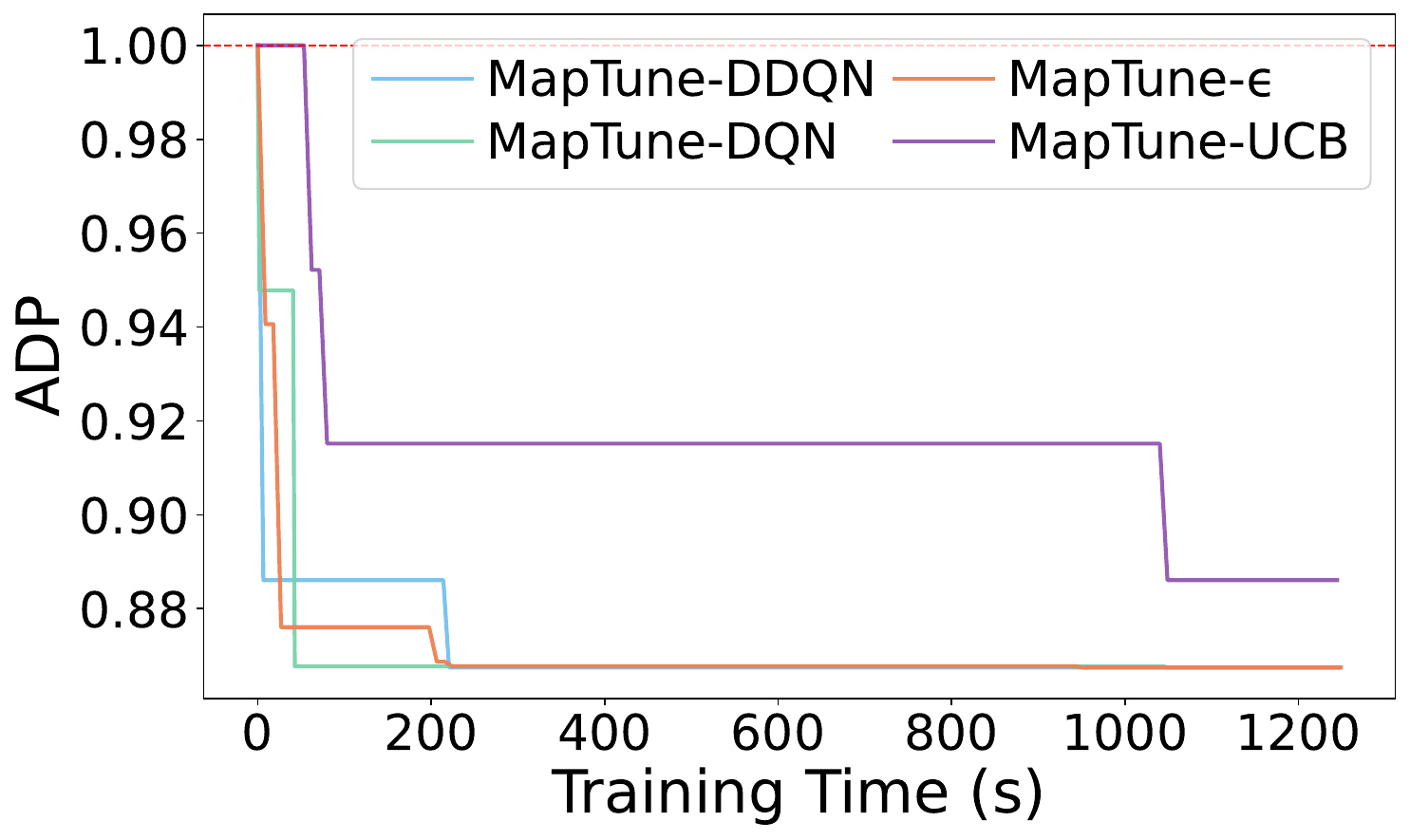}
        \caption{c1355}
        \label{fig:comp7:1355}
    \end{subfigure}
    \hfill
    \begin{subfigure}[b]{0.32\textwidth}
        \includegraphics[width=0.99\textwidth]{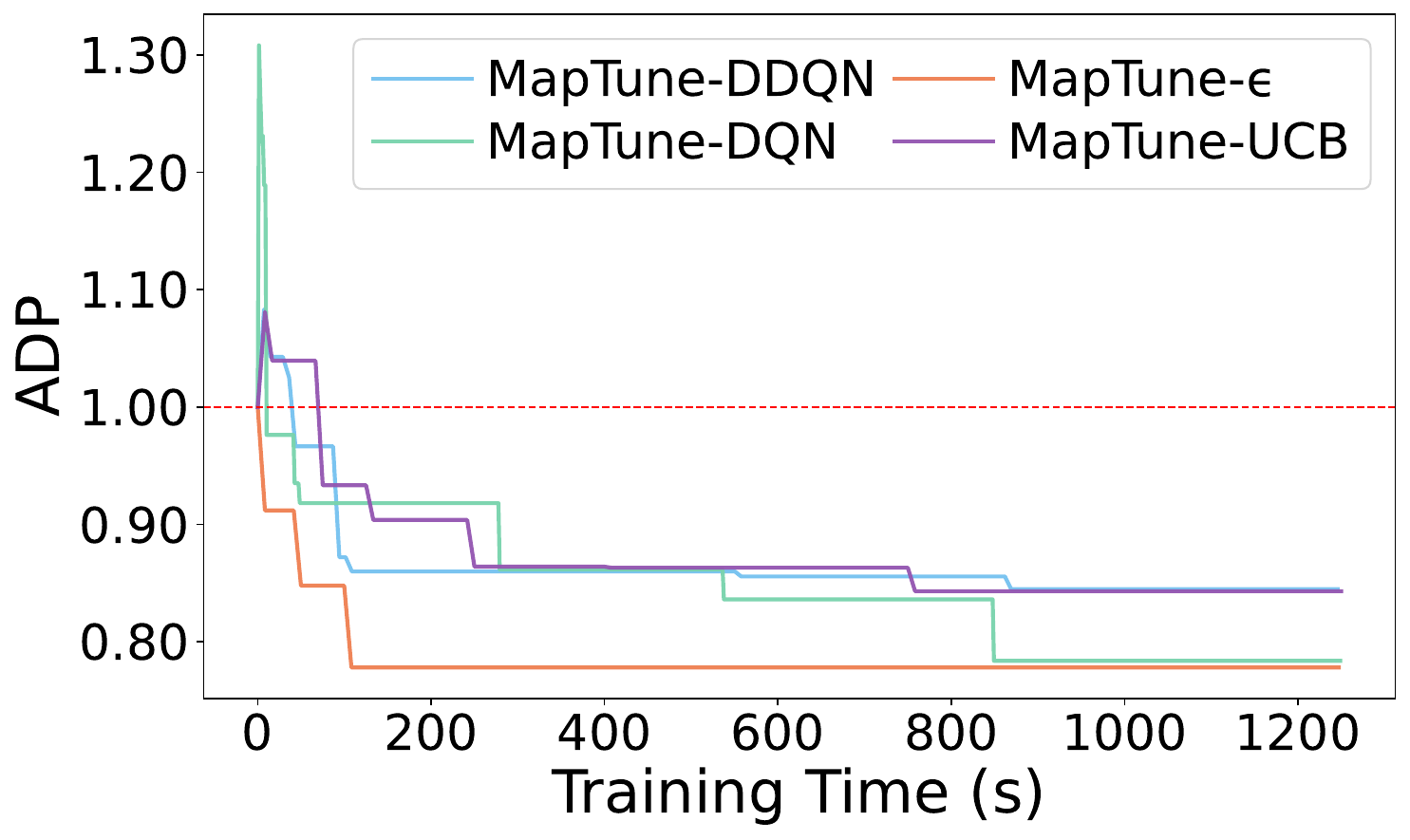}
        \caption{c880}
        \label{fig:comp7:c880}
    \end{subfigure}
    \hfill
    \begin{subfigure}[b]{0.32\textwidth}
        \includegraphics[width=0.99\textwidth]{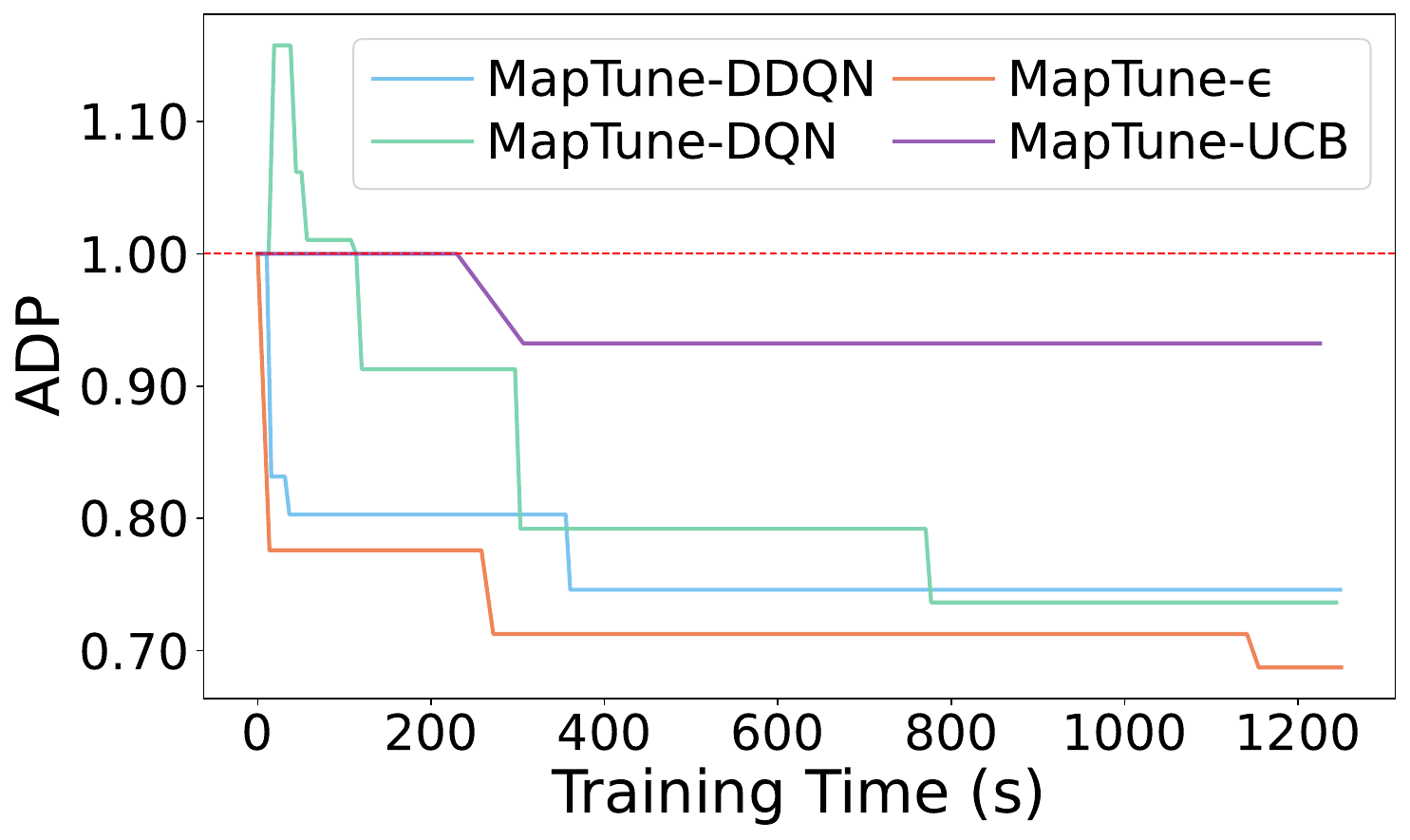}
        \caption{ode}
        \label{fig:comp7:ode}
    \end{subfigure}
    
    \medskip
    
    \begin{subfigure}[b]{0.32\textwidth}
        \includegraphics[width=0.99\textwidth]{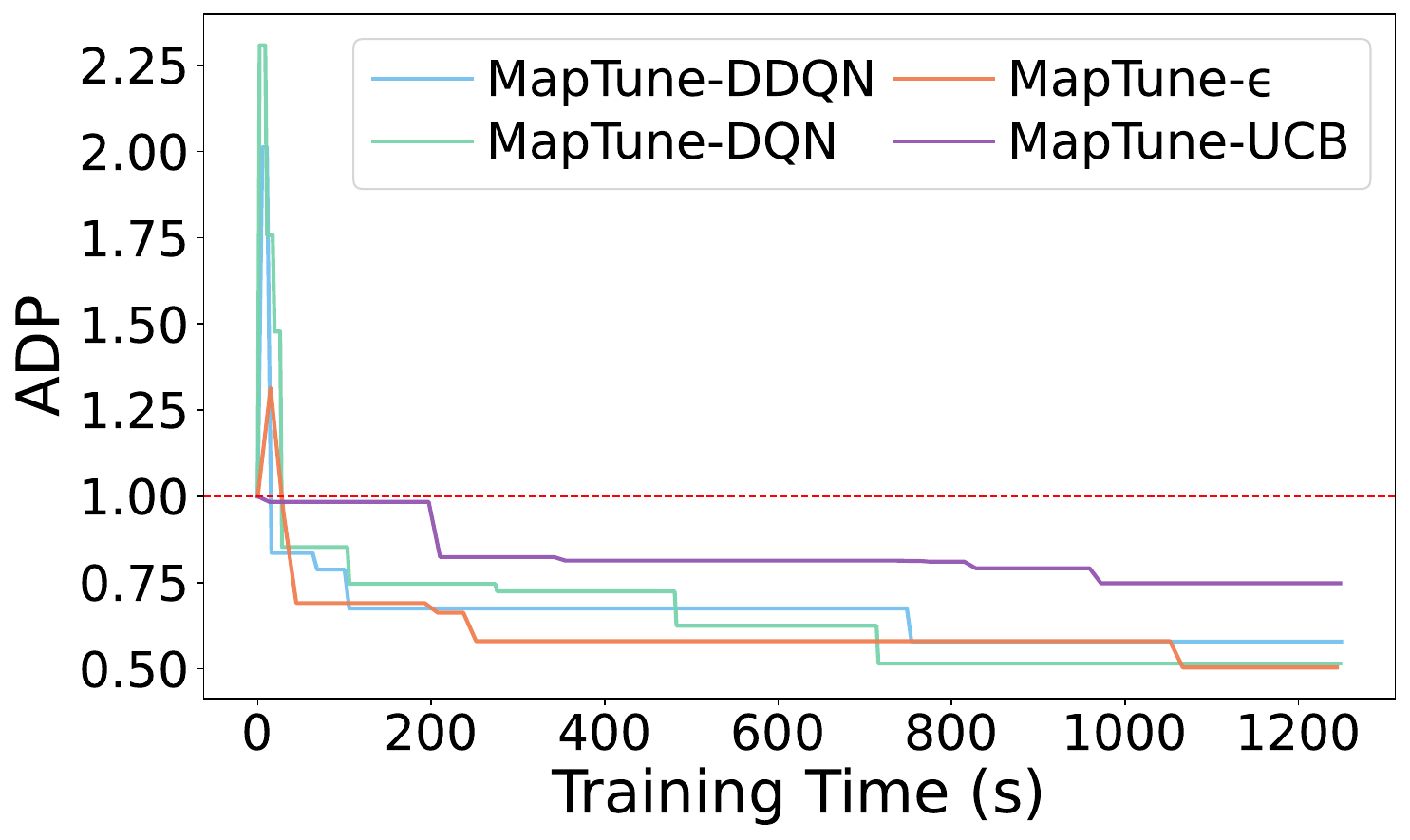}
        \caption{s35932}
        \label{fig:comp7:s35932}
    \end{subfigure}
    \hfill
    \begin{subfigure}[b]{0.32\textwidth}
        \includegraphics[width=0.99\textwidth]{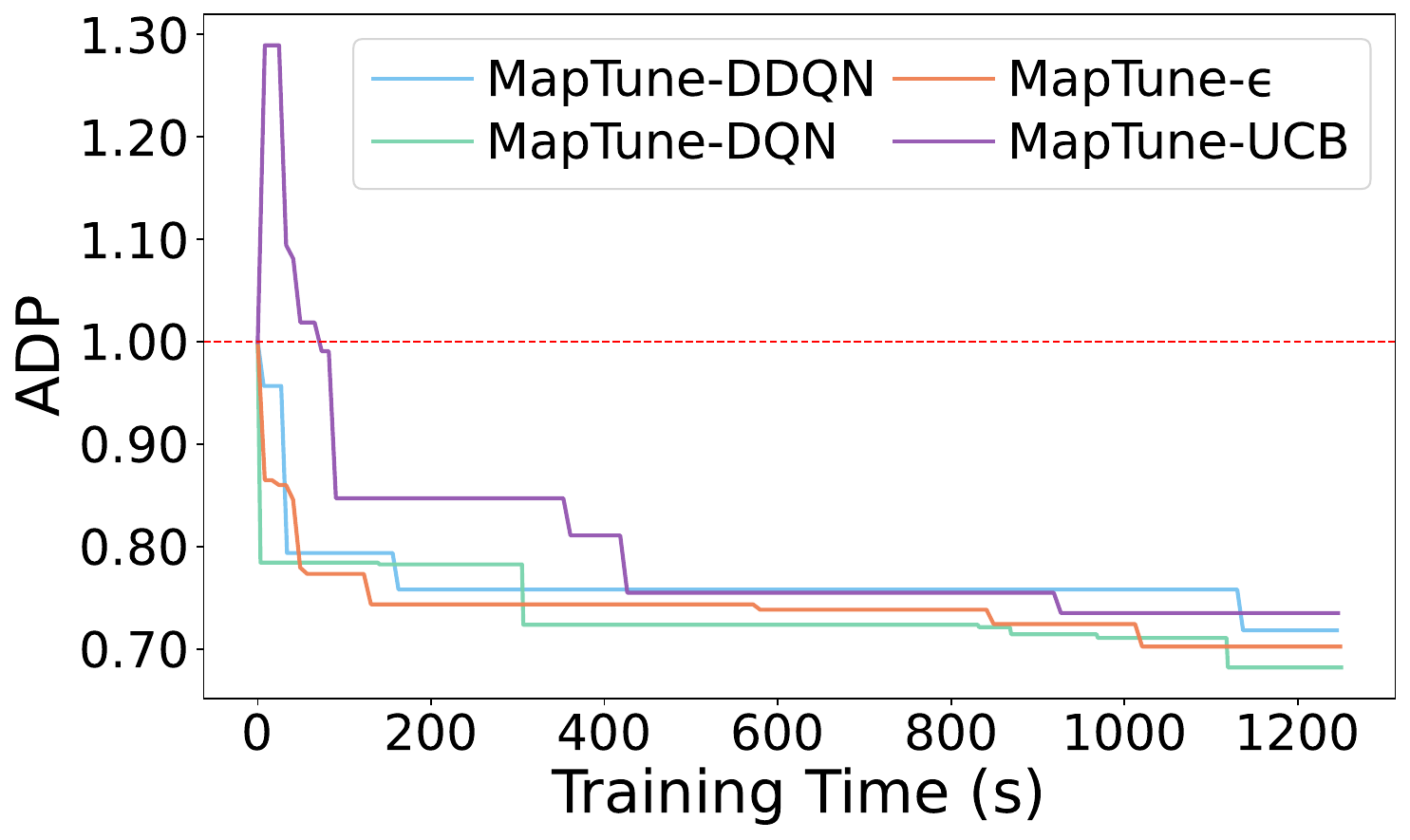}
        \caption{s838a}
        \label{fig:comp7:s838a}
    \end{subfigure}
    \hfill
    \begin{subfigure}[b]{0.32\textwidth}
        \includegraphics[width=0.99\textwidth]{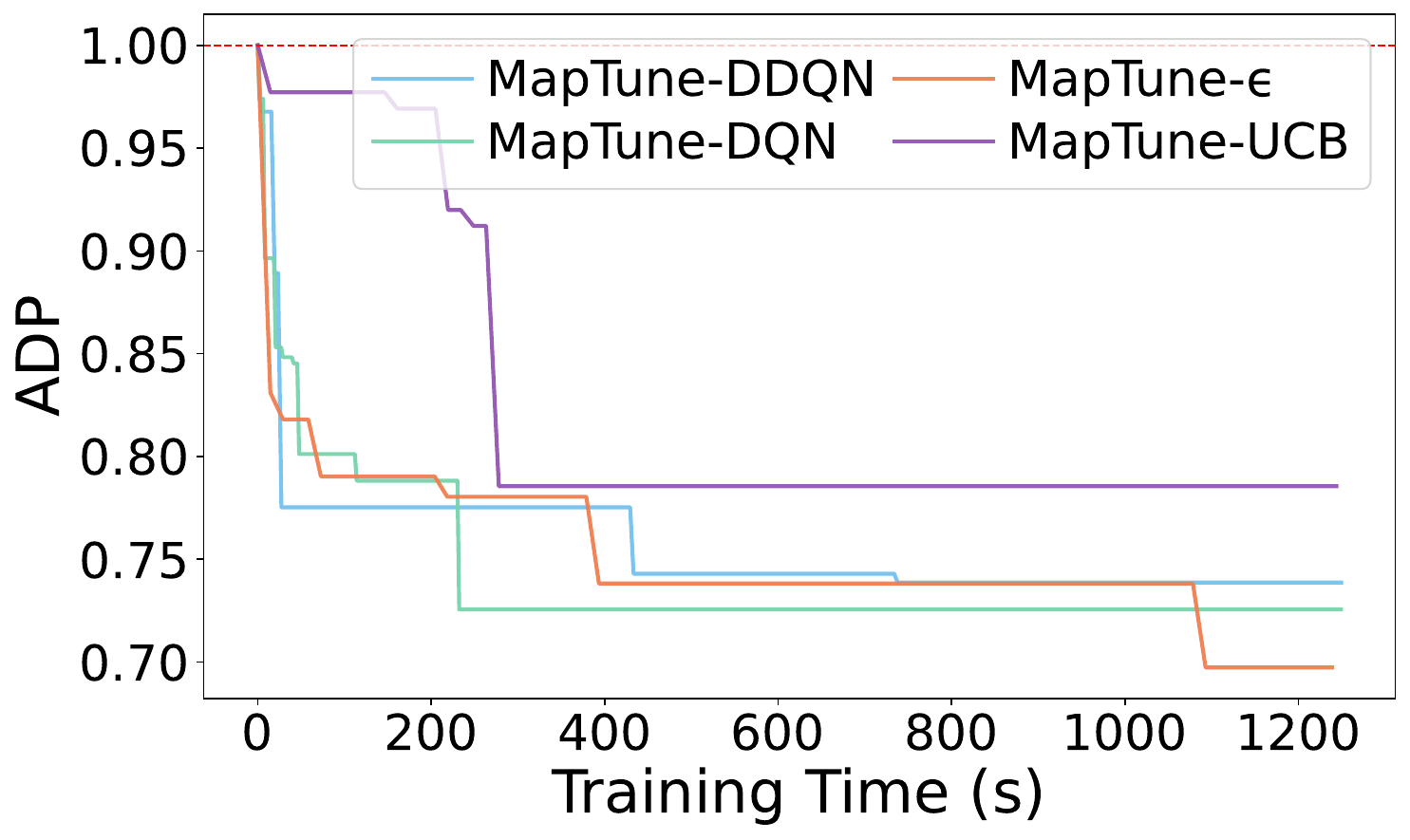}
        \caption{sin}
        \label{fig:comp7:sin}
    \end{subfigure}
    
    \caption{Comparison of ADP convergence rates of nine selected designs mapped on \texttt{ASAP7} library tuned by various MapTune-MAB and MapTune-Q methods with ABC Delay-driven mapper. Baselines (constant one) are collected with the original technology library. *The lower the better.}
    \vspace{-3mm}
    \label{fig:compare_all_7nm}
\end{figure*}

Due to the variations of the proposed approaches, we structure the experimental results to answer the three following research questions (RQ):

\vspace{1mm}
\noindent
\textbf{\underline {RQ1}: How effective is MapTune in optimizing ADP?}




\noindent
\textbf{MapTune shows a stable convergence rate regardless of methods/designs.} 
In Section \ref{section:approach}, we employ Normalized $ADP$ as a single metric to guide our MapTune framework. Here, we compare the normalized $ADP$ optimization trends across selected designs and methods. Due to page limits, we choose nine representative designs mapped on \texttt{ASAP7} library with Delay-driven mapper, as depicted in Figure \ref{fig:compare_all_7nm}. We focus on the first 1200 seconds time span to emphasize the rapid convergence rates of the different methods within MapTune.

As illustrated in Figure \ref{fig:compare_all_7nm}, across nine selected designs, various MapTune methods are able to converge to a lower achivable $ADP$ within the given optimization time span. Take design \texttt{bar} in Figure \ref{fig:comp7:bar} as an example, all four MapTune methods can achieve at least $\sim$15\% ADP reduction within 300 seconds, while MapTune-UCB can obtain an over 20\% ADP reduction in the first 15 seconds which is significantly rapid. 

Despite the rapid convergence feature among various methods and their variational settings, we do notice MapTune-MAB methods show a slight superiority than MapTune-Q methods in terms of general convergence rates and lowest achievable $ADP$. For instance, for design \texttt{c880} as shown in Figure \ref{fig:comp7:c880}, MapTune-$\epsilon$ converges to the lowest $ADP$ within 160 seconds, while MapTune-DDQN eventually achieves a $\sim$5\% higher $ADP$ despite a similar convergence time. MapTune-DQN converges to a similar $ADP$ as of MapTune-$\epsilon$ however suffers more than $5\times$ convergence time. 


\vspace{1mm}
\noindent
\textbf{\underline{RQ2}: Are MapTune adaptive to different technologies?}
\begin{figure*}[htbp]
    \centering
    \begin{subfigure}[b]{0.49\textwidth}
        \includegraphics[width=0.95\textwidth]{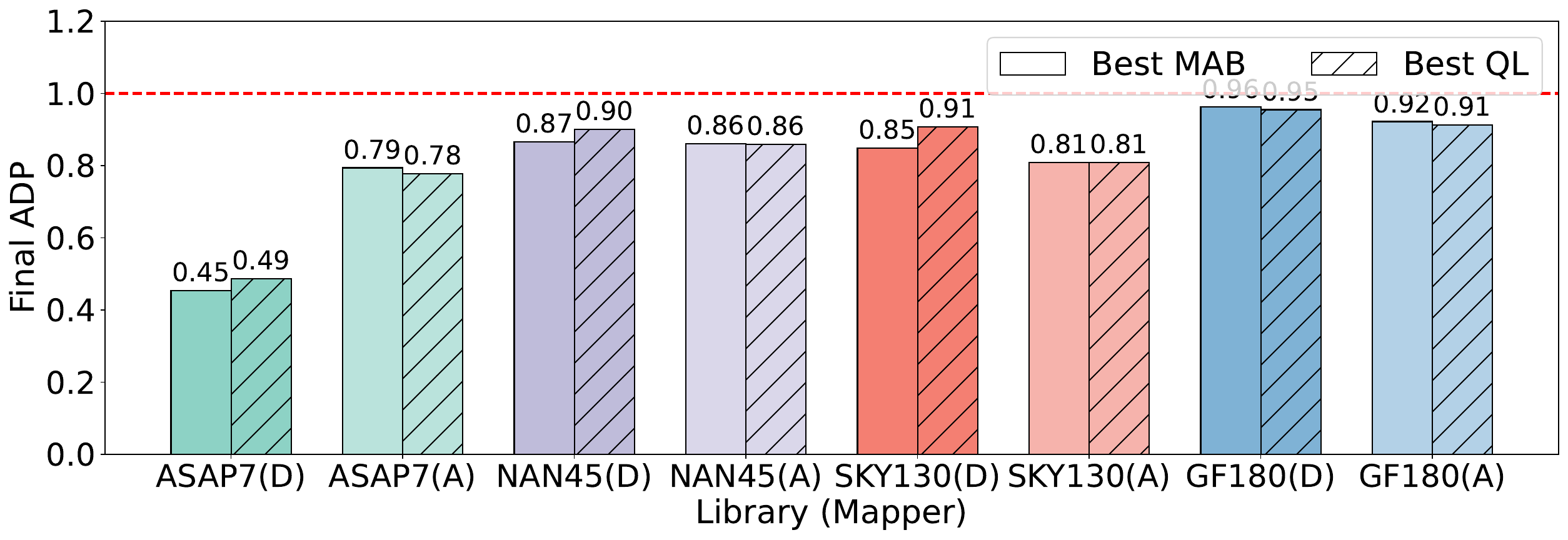}
        \caption{b14}
        \label{fig:bar_comp_bar:b14}
    \end{subfigure}
    \hfill
    \begin{subfigure}[b]{0.49\textwidth}
        \includegraphics[width=0.95\textwidth]{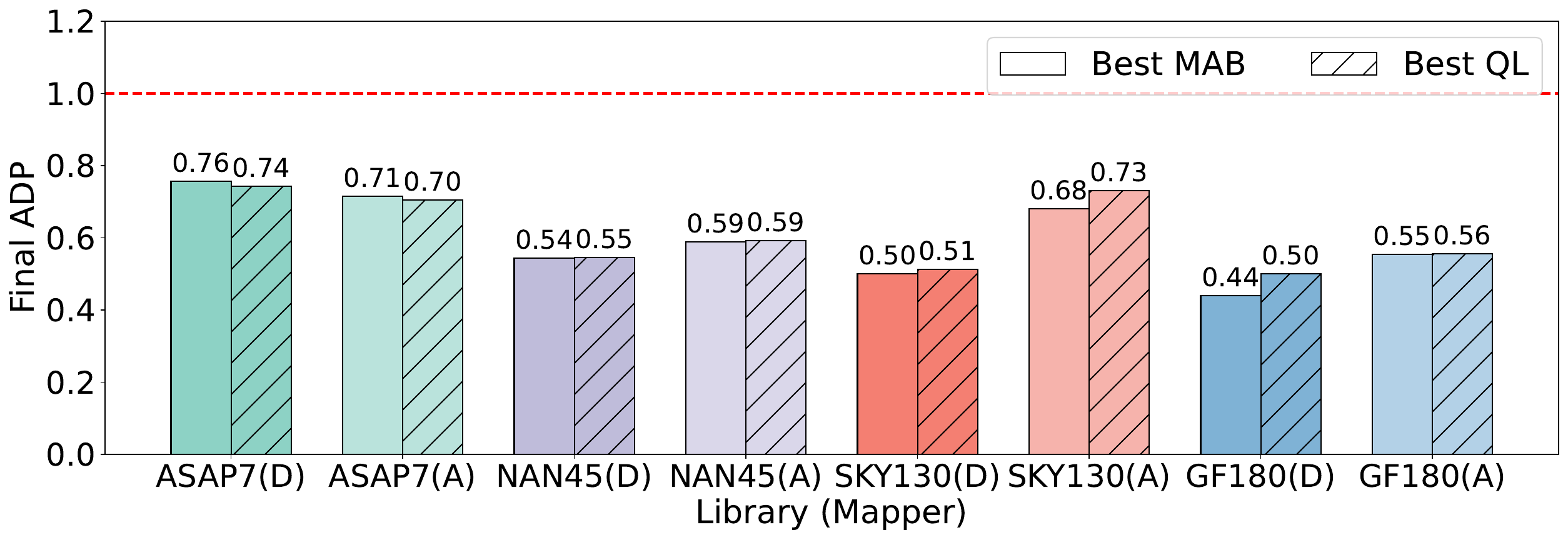}
        \caption{bar}
        \label{fig:bar_comp_bar:bar}
    \end{subfigure}
    
    \medskip
    \begin{subfigure}[b]{0.49\textwidth}
        \includegraphics[width=0.95\textwidth]{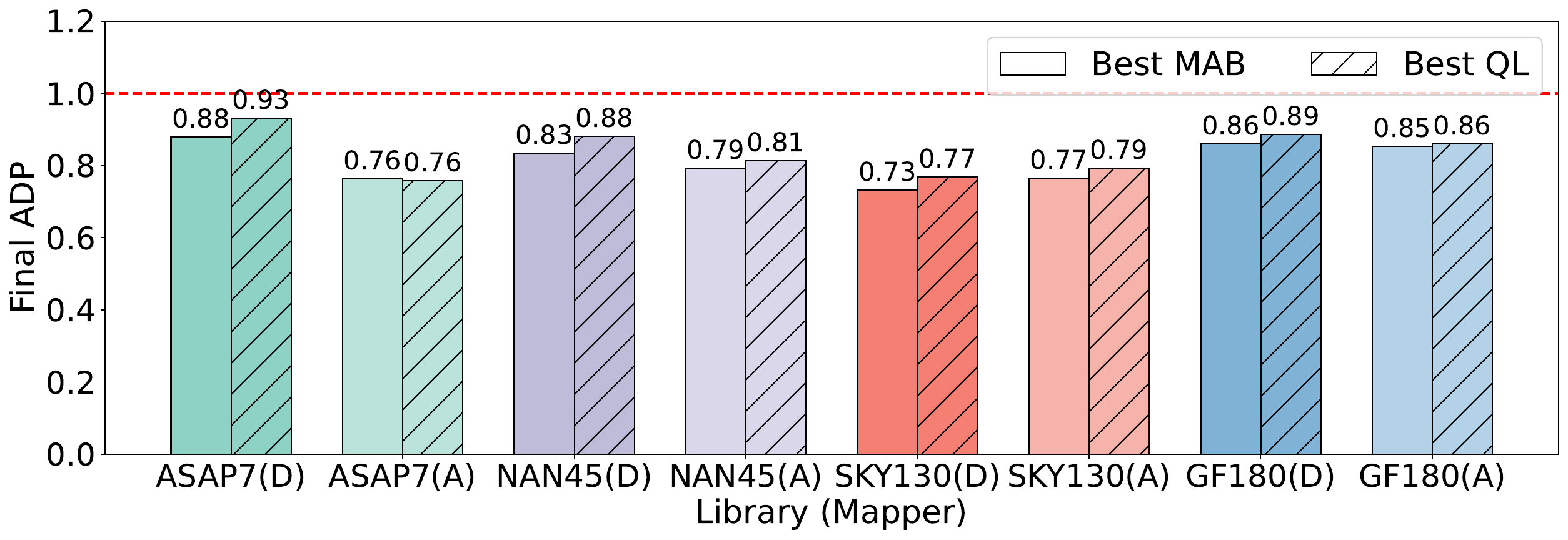}
        \caption{c1238}
        \label{fig:bar_comp_bar:c1238}
    \end{subfigure}
    \hfill
    \begin{subfigure}[b]{0.49\textwidth}
        \includegraphics[width=0.95\textwidth]{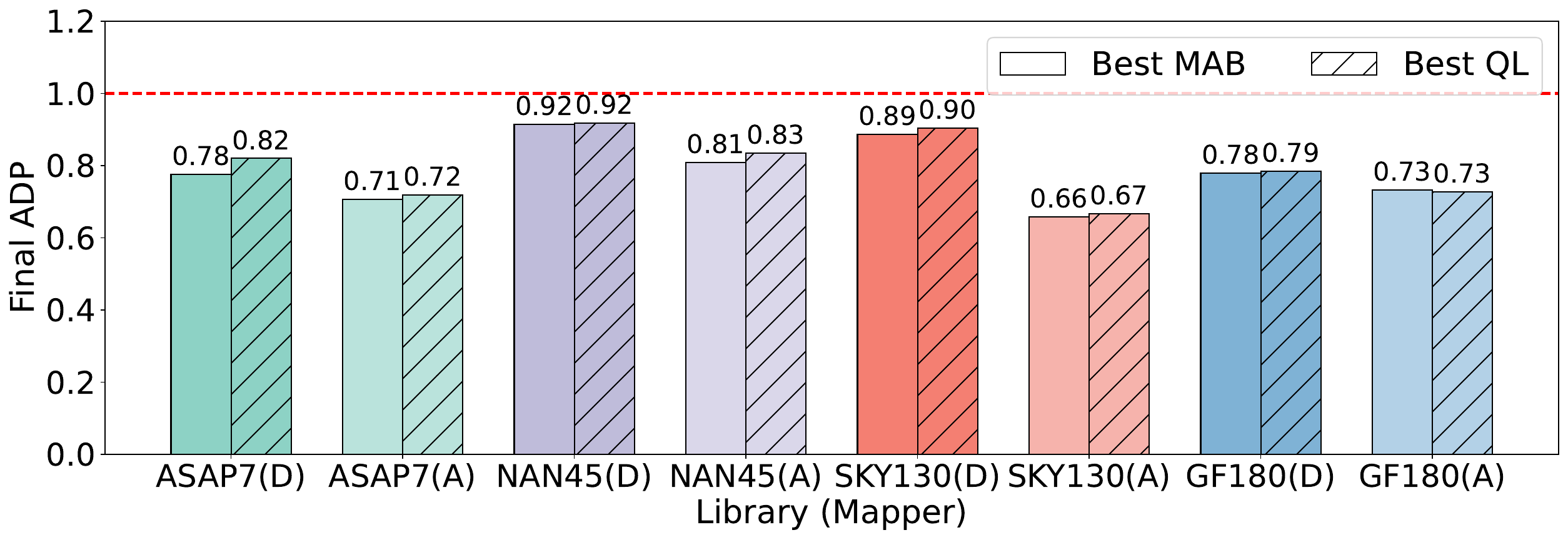}
        \caption{c880}
        \label{fig:bar_comp_bar:c880}
    \end{subfigure}
    
    \medskip
    \begin{subfigure}[b]{0.49\textwidth}
        \includegraphics[width=0.95\textwidth]{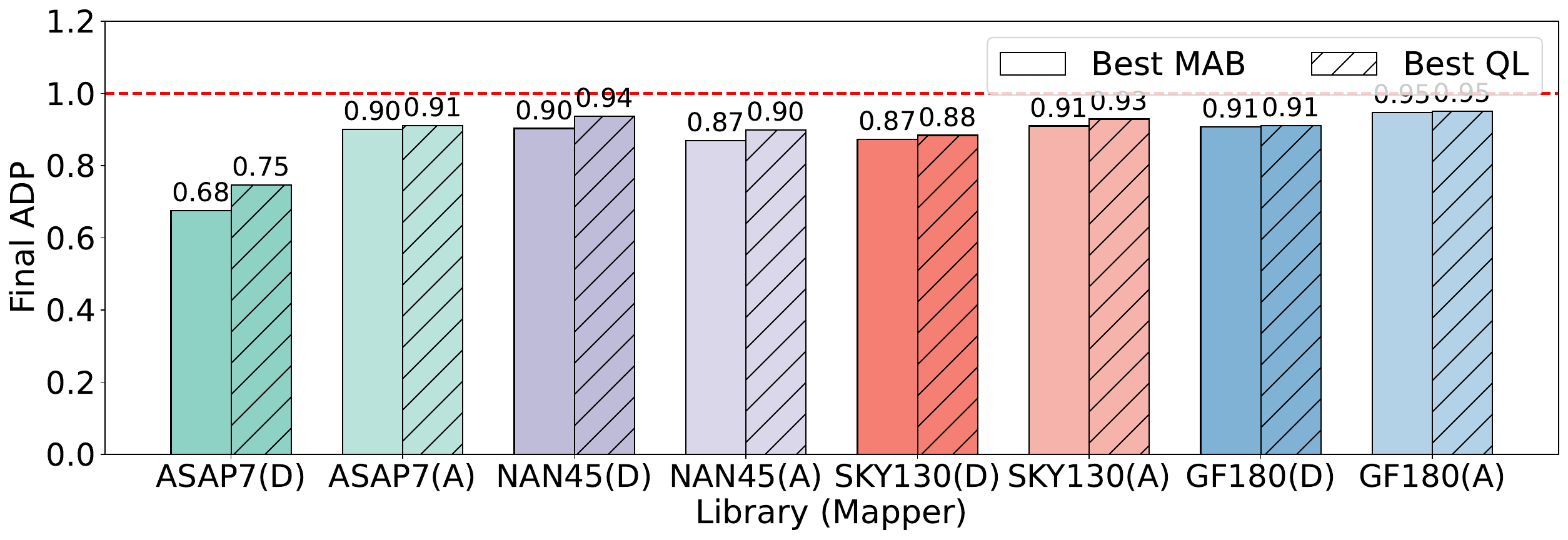}
        \caption{ode}
        \label{fig:bar_comp_bar:ode}
    \end{subfigure}
    \hfill
    \begin{subfigure}[b]{0.49\textwidth}
        \includegraphics[width=0.95\textwidth]{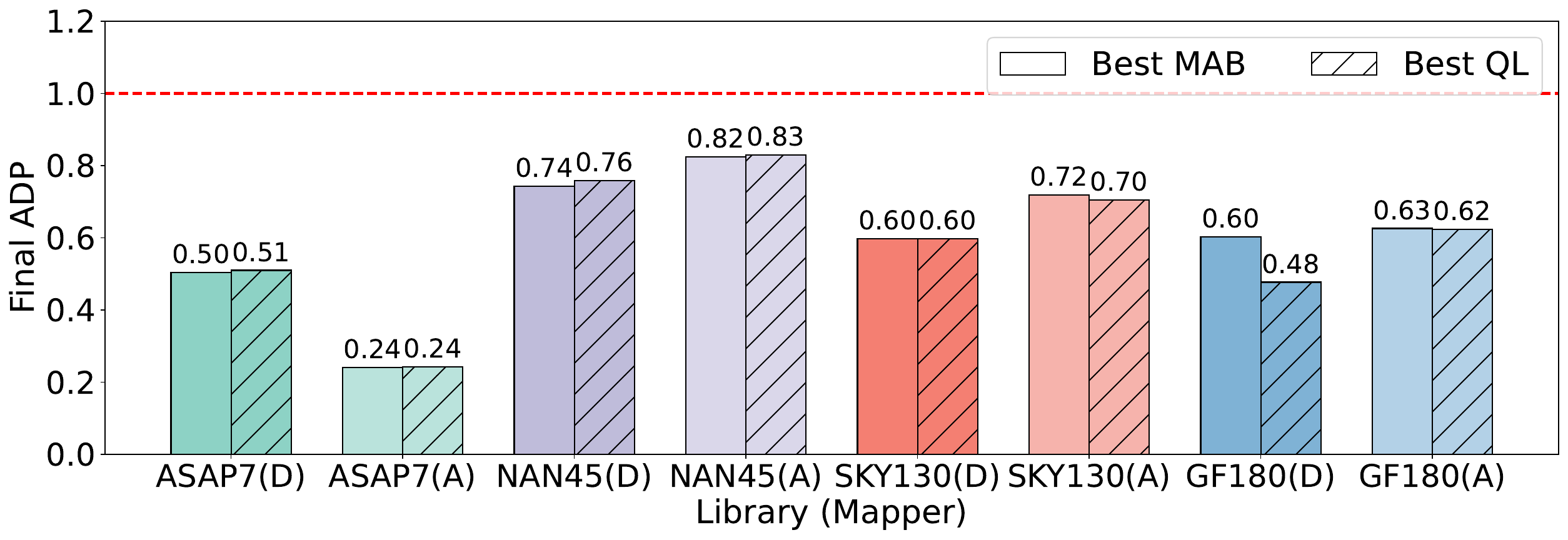}
        \caption{s35932}
        \label{fig:bar_comp_bar:s35932}
    \end{subfigure}
    
    \medskip
    \begin{subfigure}[b]{0.49\textwidth}
        \includegraphics[width=0.95\textwidth]{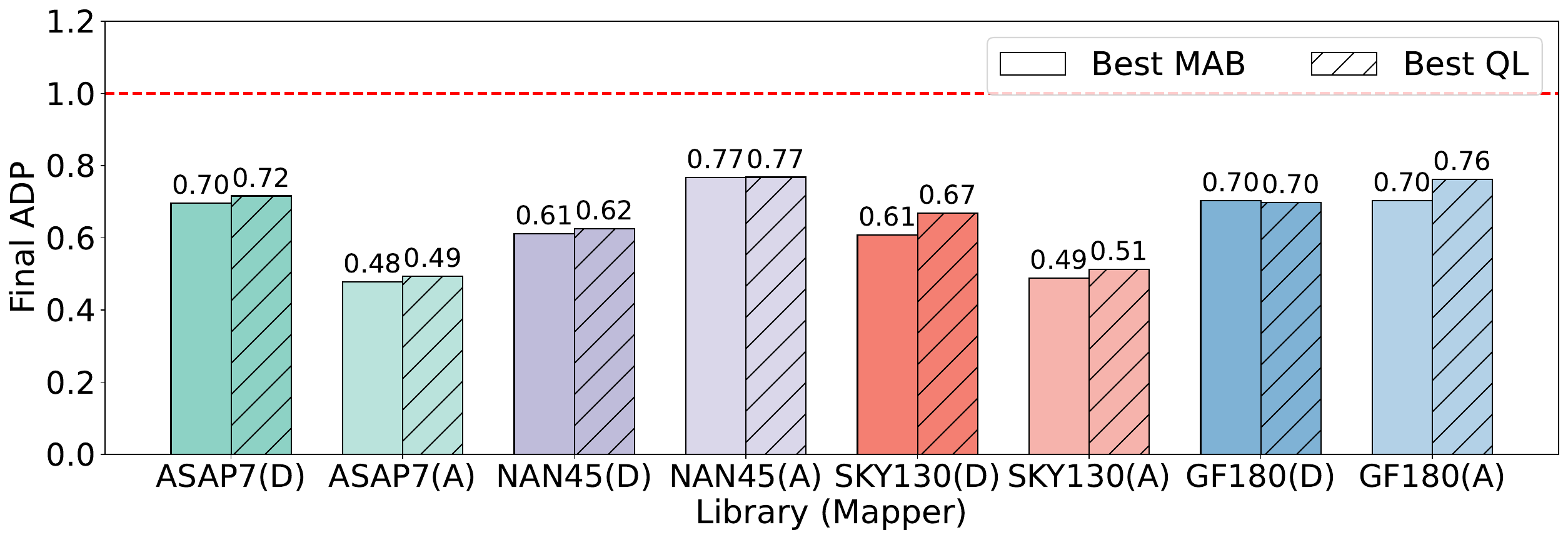}
        \caption{s838a}
        \label{fig:bar_comp_bar:s838a}
    \end{subfigure}
    \hfill
    \begin{subfigure}[b]{0.49\textwidth}
        \includegraphics[width=0.95\textwidth]{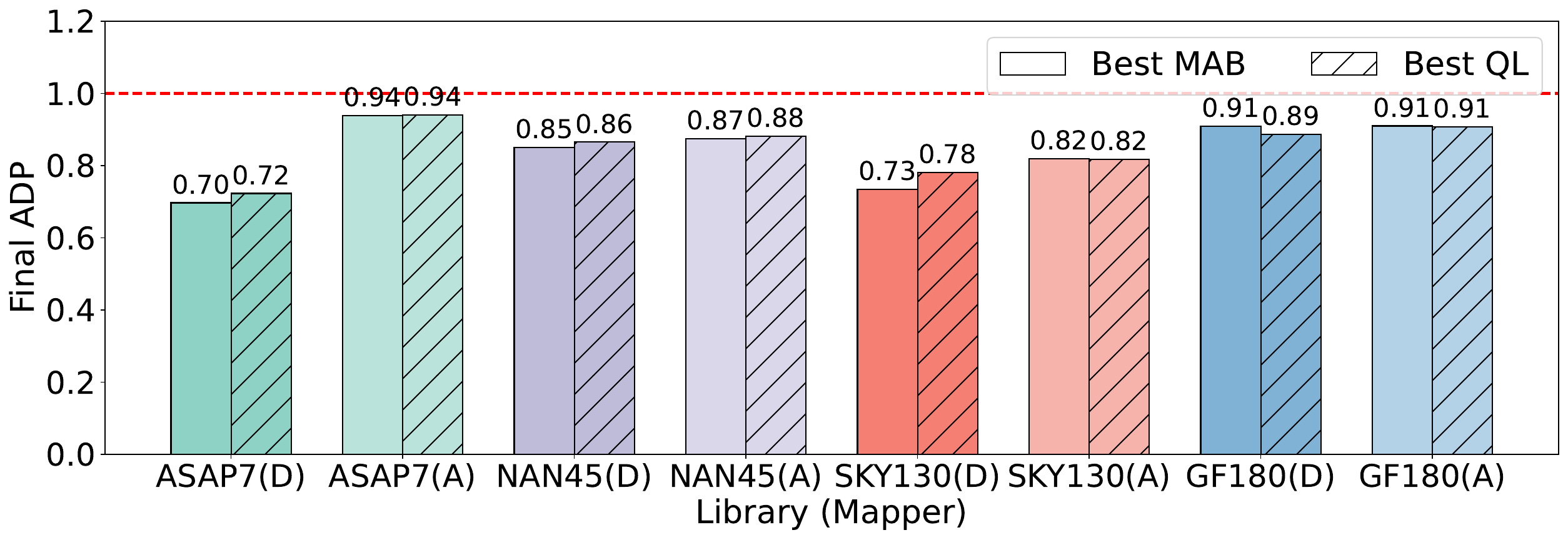}
        \caption{sin}
        \label{fig:bar_comp_bar:sin}
    \end{subfigure}
    
    \caption{Comparison of final converged ADP of eight selected designs mapped on various technology libraries tuned by MapTune-MAB and MapTune-Q methods with ABC Delay-driven (D) and Area-driven (A) mappers. Baselines (constant one) are collected with the original technology library. *The lower the better.}
    \label{fig:compare_all_bar_2_mapper}
\end{figure*}

\noindent
\textbf{MapTune is effective regardless of various technology libraries.} In Figure \ref{fig:compare_all_bar_2_mapper}, we showcase the final converged $ADP$ for eight selected designs on four libraries tuned by MapTune. Note that, in this figure, we choose the one with the best final converged $ADP$ within MapTune-MAB and MapTune-Q methods respectively, denoted as Best MAB and Best QL.

As shown in Figure \ref{fig:compare_all_bar_2_mapper}, for the same design mapped on different technologies, MapTune achieves at least a 4\% $ADP$ reduction (as shown in Figure \ref{fig:bar_comp_bar:b14}, design \texttt{b14} mapped on \texttt{GF180} tuned by MapTune-MAB with Delay-driven mapper) over the baseline (i.e., all cells retained in the original library, depicted by the red dashed line). More specifically, for design \texttt{s838a} mapped on various technology libraries tuned by MapTune, an average of 36\% $ADP$ reduction can be obtained. This highlights effectiveness of MapTune on diverse technology libraries.

\vspace{1mm}
\noindent
\textbf{\underline{RQ3}: How effective is MapTune regarding different technology mappers?}


\noindent
\textbf{MapTune is effective regardless of different technology mappers.} Besides various technology libraries are compared in Figure \ref{fig:compare_all_bar_2_mapper}, both ABC built-in Delay-driven and Area-driven mappers are also evaluated for MapTune framework. For the eight selected designs with both mappers on different technology libraries tuned by MapTune, the final converged ADP are all brought to a lower level comparing to the baseline. For example, for design \texttt{s35932} shown in Figure \ref{fig:bar_comp_bar:s35932}, MapTune achives an average of 40.12\% ADP reduction with Delay-driven mapper while an average of 40.00\% ADP reduction with Area-driven mapper highlighting that MapTune is effective regardless of technology mappers.

\vspace{-1mm}
\subsection{Pareto-Optimal Exploration}
While we have confirmed that MapTune is able to identify design point that significantly improves the quality-of-results (QoR), we want to see whether MapTune is able to identify Pareto frontier of technology mapping. One of the key aspects in this domain is the trade-off between delay and area, two primary metrics that dictate the efficiency and compactness of a given solution. With the results at hand, the focus lies on discussing how the algorithm has managed to identify a new frontier that is superior to the baseline and confirming the perpetual trade-off between delay and area.


\begin{table*}
\centering
\caption{Detailed delay/area comparison results between ABC (baseline) and MapTune in Delay-driven mapping. Our results shows more than 20.34\% delay improvements with slightly area increase (<2\%) or simultaneous area reduction, on average. }
\vspace{-3mm}
\label{tab:Delay-driven-mapper}
\resizebox{0.99\textwidth}{!}{%
\begin{tabular}{|ccccccccccccc|}
\hline
\multicolumn{13}{|c|}{\textbf{Delay ($ps$)}}                                                                                                                                                                                                                                                                                                                                                                                                                                                                                                                                                     \\ \hline
\multicolumn{1}{|c|}{\multirow{2}{*}{\textbf{Design}}} & \multicolumn{3}{c|}{\textbf{\texttt{ASAP7}}}                                                                                                             & \multicolumn{3}{c|}{\textbf{\texttt{NAN45}}}                                                                                                           & \multicolumn{3}{c|}{\textbf{\texttt{SKY130}}}                                                                                                            & \multicolumn{3}{c|}{\textbf{\texttt{GF180}}}                                                                                           \\ \cline{2-13} 
\multicolumn{1}{|c|}{}                                 & \multicolumn{1}{c|}{\textbf{Baseline}} & \multicolumn{1}{c|}{\textbf{Best MAB}}            & \multicolumn{1}{c|}{\textbf{Best RL}}            & \multicolumn{1}{c|}{\textbf{Baseline}} & \multicolumn{1}{c|}{\textbf{Best MAB}}            & \multicolumn{1}{c|}{\textbf{Best RL}}            & \multicolumn{1}{c|}{\textbf{Baseline}} & \multicolumn{1}{c|}{\textbf{Best MAB}}             & \multicolumn{1}{c|}{\textbf{Best RL}}             & \multicolumn{1}{c|}{\textbf{Baseline}} & \multicolumn{1}{c|}{\textbf{Best MAB}}              & \textbf{Best RL}              \\
\hline
\end{tabular}%
}
\end{table*}
\begin{table*}
\centering
\caption{Detailed delay/area comparison results between ABC (baseline) and MapTune in Area-driven mapping. Our results shows more than 20.85\% delay improvements with slightly area increase or simultaneous area reduction, on average.}
\vspace{-3mm}
\label{tab:Area-driven-mapper}
\resizebox{0.99\textwidth}{!}{%
\begin{tabular}{|ccccccccccccc|}
\hline
\multicolumn{13}{|c|}{\textbf{Delay ($ps$)}}                                                                                                                                                                                                                                                                                                                                                                                                                                                                                                                                                                                                            \\ \hline
\multicolumn{1}{|c|}{\multirow{2}{*}{\textbf{Design}}} & \multicolumn{3}{c|}{\textbf{\texttt{ASAP7}}}                                                                                                             & \multicolumn{3}{c|}{\textbf{\texttt{NAN45}}}                                                                                                           & \multicolumn{3}{c|}{\textbf{\texttt{SKY130}}}                                                                                                            & \multicolumn{3}{c|}{\textbf{\texttt{GF180}}}                                                                                         \\ \cline{2-13} 
\multicolumn{1}{|c|}{}                                 & \multicolumn{1}{c|}{\textbf{Baseline}} & \multicolumn{1}{c|}{\textbf{Best MAB}}            & \multicolumn{1}{c|}{\textbf{Best RL}}            & \multicolumn{1}{c|}{\textbf{Baseline}} & \multicolumn{1}{c|}{\textbf{Best MAB}}            & \multicolumn{1}{c|}{\textbf{Best RL}}              & \multicolumn{1}{c|}{\textbf{Baseline}} & \multicolumn{1}{c|}{\textbf{Best MAB}}             & \multicolumn{1}{c|}{\textbf{Best RL}}             & \multicolumn{1}{c|}{\textbf{Baseline}} & \multicolumn{1}{c|}{\textbf{Best MAB}}             & \textbf{Best RL}              \\
\hline
\end{tabular}%
}
\end{table*}

We present the exact Delay and Area results in Table \ref{tab:Delay-driven-mapper} and Table \ref{tab:Area-driven-mapper} from the Delay-driven mapper, and Area-driven mapper, respectively. 
For space constraints, we showcase results from 20 selected designs across the entire benchmark suites. The final mapping results for each method were obtained at the one-hour timeout, corresponding to the best achieved ADP for each MapTune-MAB and MapTune-Q method. 

Analyzing the tables reveals that MapTune emphasize on delay optimization with both ABC technology mappers. Across designs, libraries, and methods, we observe an average delay reduction of 21.77\%, while area reductions average only 0.79\% for Delay-driven mapper. Similarly, for Area-driven mapper, the average delay reduction is 22.30\%, but there exists an area penalty of 0.5\% on average. In fact, delay tends to benefit more from the tuned technology libraries optimized by MapTune with modest area trade-offs. This is evident in cases such as the design \texttt{multiplier} mapped with \texttt{SKY130} library tuned by MapTune-MAB which achieves a substantial 39.96\% delay improvement with a 4.03\% area penalty as shown in Table \ref{tab:Delay-driven-mapper}. There still exists opposite cases that indicating trade-off delay for area optimization. As for design \texttt{sqrt} mapped on \texttt{GF180} library tuned by MapTune-MAB with Area-driven Mapper shown in Table \ref{tab:Area-driven-mapper}, MapTune introduces 1.25\% delay penalty resulting a 14.90\% area reduction ultimately. 
Such Pareto-Optimal trade-offs are often desirable, as significant delay/area reductions can potentially benefits more throughout the design flow than the introduced modest area/delay penalty.

\section{Conclusions}

This paper explores the use of partially sampled technology libraries to reduce the search space for better QoR in technology mapping. Our case study empirically demonstrates the importance of this process, given its potential impact on the trade-off between area and delay, as well as its capability to reveal new performance Pareto frontiers. In response to this challenge, we introduce MapTune, a novel sampling framework based on Reinforcement Learning by leveraging both MAB and Q-Learning and seamlessly integrated within the ABC framework. Extensive evaluations using five distinct benchmark suites confirmed the effectiveness of MapTune framework for technology library tuning. By solely focusing on library optimization, MapTune is able to achieve an average ADP improvement of 22.54\% and identify pareto-optimal results. Future work will concentrate on cross-design library exploration and integration with automatic library generation tools.


\section*{Acknowledgement}
This work is funded by the National Science Foundation (NSF) under awards NSF 2229562, 2349670, 2349461, 2403134, and University of Maryland.

\bibliographystyle{unsrt}
\bibliography{tex/synthesis}





\end{document}